\documentclass[11pt]{article}
\usepackage[margin=18mm]{geometry}
\usepackage[superscript,biblabel]{cite}
\usepackage{authblk}

\usepackage{amsmath,amssymb}
\usepackage{siunitx}
\usepackage{upgreek}
\usepackage{tcolorbox}
\usepackage{multirow}
\usepackage{booktabs}
\usepackage{xcolor}
\definecolor{blan}{RGB}{0,128,255} 

\usepackage{hyperref}
\hypersetup{
    colorlinks = true,
    citecolor={blue},
    linkcolor={red},
    allbordercolors = {white},
    urlcolor={blue}
}

\newcommand{\Tr}{\mathrm{Tr}}

\newcommand{\lproj} [2]{\left|{#2}\right\rangle_{\!{#1}}\!\!\left\langle{#2}\right|}
\newcommand{\norm}  [1]{\left|\left|{#1}\right|\right|}
\newcommand{\avg}   [1]{\left\langle{#1}\right\rangle}

\newcommand{\Tx}{\text{Tx}}
\newcommand{\Rx}{\text{Rx}}

\newcommand{\sat}{\text{sat}}

\newcommand{\cK}{\mathcal{K}}
\newcommand{\upA}{\mathrm{A}}
\newcommand{\upB}{\mathrm{B}}
\newcommand{\upE}{\mathrm{E}}
\newcommand{\upX}{\mathrm{X}}
\newcommand{\upZ}{\mathrm{Z}}

\title{Assessment of practical satellite quantum key distribution architectures for current and near-future missions}

\author[1]{Davide Orsucci*}
\author[1]{Philipp Kleinpa\ss}
\author[2]{Jaspar Meister}
\author[1]{Innocenzo De Marco}
\author[1]{Stefanie H\"ausler}
\author[1]{Thomas Strang}
\author[3]{Nino Walenta}
\author[1]{Florian Moll}

\affil[1]{Institute of Communications and Navigation, German Aerospace Center (DLR), \newline Münchener Str. 20, 82234 Weßling, Germany}
\affil[2]{Institute for Satellite Geodesy and Inertial Sensing, German Aerospace Center (DLR),\newline Callinstr. 30b, 30167 Hannover, Germany}
\affil[3]{Fraunhofer, Heinrich Hertz Institute (HHI), Einsteinufer 37, 10587 Berlin, Germany}

\date{\today}

\begin{document}

\maketitle

\abstract{
Quantum key distribution (QKD) allows the generation of cryptographic keys beyond the computational hardness paradigm and is befitting for secure data transmission requiring long-term security. The communication distance of fibre-based QKD, however, is limited to a few hundred kilometers due to the exponential scaling of signal attenuation. Satellite QKD (SatQKD) can instead leverage free-space optical links to establish long-range connections and enable global-scale QKD. In this work we review the manifold of design choices that concur to form the set of possible SatQKD architectures. These include the choice of the QKD protocol and its physical implementation, but also the satellite orbit, the optical link direction, and whether or not to use trusted-node relays. The possible SatQKD architectures are then evaluated in terms of key generation throughput, latency and maximum reachable communication distance, but also the  system-level security and implementation complexity. Given the technical challenges of realising SatQKD systems it is paramount, for near-future satellite missions, to adhere to the simplest possible architecture that still allows to deliver the QKD service. We thus identify as advisable options the use of low-Earth orbit satellites as trusted nodes for prepare-and-measure discrete-variable QKD downlinks with weak laser pulses. The decoy-state version of BB84 is found to be the most promising QKD protocols due to the maturity of the security proofs, the high key generation rate and low system complexity. These findings are confirmed by the multitude of current and planned SatQKD missions that are adopting these architectural choices. 
}

\renewcommand\thefootnote{}
\footnotetext{
* davide.orsucci@dlr.de\\
    \textbf{Abbreviations:}
    AES,    American Encryption Standard;
    AO,     Adaptive Optics;
    BB84,   Bennett and Brassard, 1984;
    BBM92,  Bennett, Brassard, and Mermin, 1992;
    BGL,    BackGround-Light,
    BSI,    Bundesamtes für Sicherheit in der Informationstechnik, 
    COW,    Coherent One-Way; 
    CV,     Continuous Variable;
    DCR,	Dark-Count Rate;
    E91,    Ekert, 1991;
    EB,     Entanglement Based;
    DL,     DownLink;
    DLR,    Deutsches Zentrum für Luft- und Raumfahrt;
    DoS,    Denial-of-Service;
    DPS,    Differential Phase-Shift;
    DV,     Discrete Variable,
    FSO,	Free-Space Optical;
    GEO,	GEOstationary orbit;
    KMS,	Key Management System;
    LCT,    Laser Communication Terminal;
    LEO,	Low-Earth Orbit;
    LO, 	Local Oscillator;
    MDI,    Measurement-Device Independent;
    MEO,    Medium-Earth Orbit;
    MMF,    Multi-Mode Fibre;
    MP,     Mode-Pairing;
    NIST,   National Institute of Standards and Technology;
    OGS,    Optical Ground Station;
    PAA,    Point-Ahead Angle;
    PAT,    Pointing, Acquisition, and Tracking;
    PLOB,   Pirandola, Laurenza, Ottaviani, and Banchi;
    PM,     Prepare-and-Measure;
    PQC,    Post-Quantum Cryptography;
    QD,     Quantum Dot;
    QKD,    Quantum Key Distribution;
    SARG04, Scarani, Acín, Ribordy, and Gisin, 2004;
    SatQKD, Satellite-based Quantum Key Distribution;
    SKL,    Secure Key Length; 
    SKR,    Secure Key Rate;
    SMF,    Single-Mode Fibre;
    SNSPD,  Superconducting-Nanowire Single-Photon Detector;
    SPAD,   Single-Photon Avalanche Photodiode;
    SPS,    Single-Photon Source;
    SSO,    Sun-Synchronous Orbit;
    SWaP,   Size, Weight and Power;
    TF,     Twin-Field;
    TN,     Trusted Node;
    TRL,    Technology Readiness Level;
    UL,     UpLink;
    UTN,    UnTrusted Node;
    VLEO,   Very-Low-Earth Orbit;
    WCP,    Weak Coherent Pulse.
}

\renewcommand\thefootnote{\fnsymbol{footnote}}
\setcounter{footnote}{1}

\section{Introduction}
\label{sec:intro}

The modern digital communication infrastructure hinges upon public-key cryptography, which is pervasively employed to encrypt and authenticate messages. However, a full-fledged quantum computer would allow solving integer factorisation and discrete logarithms in polynomial time\cite{shor1994algorithms, proos2003shor} and thus break the security of all the standard public-key cryptographic algorithms employed today. Furthermore, the store-now decrypt-later attack amplifies this threat: an adversary could tap into a communication line, collect all the encrypted data that is transmitted across the line and then store it until the means for decrypting it become available. Given the steady progress in quantum computing hardware and given that certain highly-sensitive information is required to be maintained secret for several decades, it is of paramount importance to start addressing this threat today\cite{mosca2018cybersecurity}.

This threat can be mitigated through the use of Post-Quantum Cryptography (PQC), which aims at replacing the cryptographic primitives that are known to be broken soon by quantum algorithms, with other primitives, such as lattice-based or code-based cryptography, that are supposedly resistant to quantum computing attacks\cite{Bernstein2017}. Being based on classical encryption algorithms, PQC can be largely realised with software and firmware updates to the existing digital communication infrastructure -- the main deployment issues of PQC stem from increased key size and computation complexity, thus hardware updates will be required in applications requiring low latency or low bandwidth usage. The downside is that PQC security is still based on computational assumptions and it is entirely possible that new algorithms may in the future break some PQC primitives. Indeed, the Rainbow and SIKE cryptosystems, which were among the semi-finalists in the National Institute of Standards and Technology (NIST) evaluation for adoption as PQC standards, were broken by classical cryptoanalysis algorithms\cite{beullens2022breaking, castryck2023efficient}. This casts significant doubts on the long-term reliability of PQC encryption. Considering furthermore that quantum cryptoanalysis algorithms are under-explored and could be the origin of many new attack methods\cite{biasse2023quantum}, it may be unwise to rely on PQC alone, especially for high-security applications.

A new approach to security came to light when Bennett and Brassard in 1984 (BB84)\cite{bennett1984quantum} introduced the first QKD protocol. Thereby quantum principles are employed to achieve security, rather than relying on computational assumptions to prove security and it can be mathematically guaranteed that even a computationally unbounded adversary cannot extract any information about the QKD key\cite{renner2008security}. Therefore, using QKD for the generation of secure keys (both for pre-quantum and post-quantum algorithms) is particularly suited in applications requiring high levels of security or long-term storage of encrypted or signed information. The main downside of QKD is that the maximum communication range through optical fibres is limited to a few hundred kilometres, even under controlled experimental conditions\cite{wang2022twin}. This stems from the fact that quantum information cannot be copied nor amplified; therefore, the intensity of a quantum signal decays exponentially along an optical fibre and is eventually overcome by noise. Quantum repeaters may in the future allow the efficient relay of quantum information, but the enabling technologies may require decades to reach a sufficient level of maturity\cite{azuma2023quantum}. 
Alternatively, long-distance quantum links could be established using currently available technology by employing Free-Space Optical (FSO) communication over satellite links. The main advantage of Satellite-based QKD (SatQKD) is that in satellite links the signal intensity loss is mainly due to beam divergence, which results in a transmission efficiency that decreases quadratically with the distance, rather than exponentially as in fibre-based systems.

In this paper, we aim to assess the strengths and weaknesses of general SatQKD architectures. In Section~\ref{sec:overview} we provide some general background information on QKD. In Section~\ref{sec:security} we discuss security aspects of QKD, with focus on aspects specific to satellite realisations. In Section~\ref{sec:status} we provide an overview of the current SatQKD literature and activities.
In Section~\ref{sec:architecture} we identify a reference SatQKD architecture and we thoroughly motivate the design choices leading to its selection. In Section~\ref{sec:comparison} we then quantitatively compare the performance of some selected SatQKD implementations, employing the previously identified architecture. In Section~\ref{sec:conclusions} we give our conclusions and an outlook for further investigations.

\section{High-level overview of quantum key distribution}
\label{sec:overview}

In this section we establish the terminology and provide a high-level overview of QKD, remarking that the topic has too many theoretical, experimental and technological facets, to be comprehensively covered here. Furthermore, several excellent reviews of QKD have already been written\cite{scarani2009security, pirandola2020advances, xu2020secure} and we refer the reader to these reviews for more thorough introductions to the subject.

\subsection{Definition of the QKD functionality}

The goal of QKD is to establish a cryptographic key (i.e., a secret bit-string) between a pair of end-users, customarily denoted as Alice and Bob, secured against a third party, usually called Eve. The key is considered secure if it is private, integer and authentic: privacy means that no third party (an eavesdropper) can have any information about it; integrity means that no third party (a tamperer) can modify the key without being detected by the legitimate communication parties; authentic means that the identity of the sender is verified, e.g., through the use of a digital signature.
Availability of the key, however, cannot be guaranteed, as it is assumed that Eve may have complete control over the quantum channel and, thus, she is in theory allowed to perform Denial-of-Service (DoS) attacks.

From an abstract-cryptography perspective, QKD is a protocol which uses a quantum communication channel (in practice, an optical link, either fibre-based or in free-space) together with a classical authenticated channel as communication resources.
Both, provided that the procedure does not abort, are used to distribute a secure key to Alice and Bob. The security is \textit{information-theoretic}, so that it holds even against a computationally unbounded adversary, and \textit{composable}, so that the key distribution functionality can be securely employed in arbitrary cryptographic scenarios\cite{ben2005universal, portmann2022security}. A natural application is to use it in conjunction with One-Time-Pad\footnote{In SatQKD the amount of generated key is expected to be rather small, therefore it may be more practical to combine them with strong symmetric encryption schemes, such as the American Encryption Standard (AES), which only require a short key to encrypt arbitrarily long messages and which are believed to be secure against quantum computer attacks.} (OTP), providing information-theoretic encryption\cite{shannon1949communication}, and with the Wegman-Carter scheme, providing information-theoretic authentication\cite{wegman1981new}. But one may also use a QKD key in arbitrary encryption schemes, e.g., to securely relay a key from one node to the next, as will be discussed later.

In more detail, the security of the key is quantified by a correctness parameter $\varepsilon_\text{corr}$ and a secrecy parameter $\varepsilon_\text{sec}$. The probability that Alice's key $k_\upA$ and Bob's key $k_\upB$ are equal ($k\triangleq k_\upA=k_\upB$) is at least $1-\varepsilon_\text{corr}$ and, furthermore, the probability that the protocol does not abort and the key is not completely uniformly random for the eavesdropper is at most $\varepsilon_\text{sec}$. The global security parameter is then given as $\varepsilon = \varepsilon_\text{sec} + \varepsilon_\text{corr}$. In quantum information language, the global security can also be expressed as follows\cite{portmann2022security}. At the end of the protocol Alice and Bob will share classical information that can be correlated with Eve's quantum information, which is described as \textit{classical-quantum state}. The \textit{ideal} QKD functionality (i.e., the mathematical abstraction of perfect QKD) distributes to Alice, Bob, and Eve the following state
\begin{align}
    \rho_{\upA\upB\upE}^\text{ideal} =  
    p_\perp \lproj{\upA}{\perp} \otimes \lproj{\upB}{\perp} \otimes \rho_\upE^\perp 
    \;+\;
    \frac{1-p_\perp}{2^n}\! \sum_{k \in \{0,1\}^n} \lproj{\upA}{k} \otimes \lproj{\upB}{k} \otimes \rho_\upE
\end{align}
where $\perp$ is a symbol denoting protocol abortion, occurring with probability $p_\perp$, and otherwise Alice and Bob receive a uniformly random $n$-bit key and Eve's (quantum) information $\rho_\upE$ is uncorrelated to it. The classical-quantum state summarising the information shared by Alice, Bob, and Eve at the end of the \textit{real} QKD protocol in general can be written as
\begin{align}
    \rho_{\upA\upB\upE}^\text{real} =  
    \sum_{k \in \cK} p_{k_\upA,k_\upB} \lproj{\upA}{k_\upA} \otimes \lproj{\upB}{k_\upB} \otimes \rho_\upE^{k_\upA, k_\upB}
    \qquad
    \cK \triangleq \{0,1\}^n \, \cup \perp
\end{align}
where $p_{k_\upA,k_\upB}$ is the probability that Alice and Bob obtain the outcomes $(k_\upA,k_\upB) \in \cK\times\cK$. The real protocol is then $(1-\varepsilon)$-secure if there exists a $p_\perp$ such that $\rho_{ABE}^\text{real}$ and $\rho_{ABE}^\text{ideal}$ are $\varepsilon$-indistinguishable, that is, if these states are $\varepsilon$-close in trace norm:
\begin{align}
    \norm{\rho_{\upA\upB\upE}^\text{real} - \rho_{\upA\upB\upE}^\text{ideal}}_\Tr \leq \varepsilon \,.
\end{align}

Finally, a QKD protocol that always aborts would be (trivially) secure according to these definitions. Therefore, one has to strengthen the definition and introduce a notion of \textit{robustness}. A $\delta$-robust QKD protocol can succeed in establishing a secure key with high probability, provided that the Quantum Bit Error Rate (QBER) is smaller or equal to $\delta$. The (asymptotic) Secure Key Rate (SKR) is defined as the ratio of the Secure Key Length (SKL) over the total number of uses of the quantum channel when these tend to infinity, while the security parameter goes to zero:
\begin{align}
    \text{SKR} = \lim_{\varepsilon \rightarrow 0} \lim_{N \rightarrow \infty} \frac{\text{SKL}(N, \varepsilon)}{N} .
\end{align}
The SKR is typically a monotonically decreasing function of the QBER and of the channel transmission, $\eta$. Devising protocols that have a high SKR in the presence of noise and photon transmission losses is paramount to enabling the practical application of QKD.

\subsection{Blueprint for a generic QKD protocol}

In broad strokes, a QKD protocol consists in the consecutive steps detailed below. The first step is the quantum part of the protocol and the rest consists in classical post-processing, which requires bi-directional classical communication. Some of the post-processing information may be encrypted and all the classical information must be authenticated; both tasks can be accomplished by using pre-shared randomness. This could originate from previous QKD rounds\cite{fung2010practical} or as a first initialization at system commissioning stage. 

\begin{tcolorbox}{
\begin{enumerate}
\item \textit{Quantum communication:} Alice and Bob exchange a sequence of $N$ signals over the quantum channel (potentially involving an untrusted third party). All quantum signals are immediately measured upon being received, resulting in $N_d \leq N$ successful detection events. The state preparation and measurement settings are chosen stochastically (e.g., employing a quantum random number generator\cite{ma2016quantum}) and this information, together with the measurement outcomes, is stored locally by Alice and Bob.

\item \textit{Raw key extraction (a.k.a., key sifting):} Alice and Bob use the authentic channel to announce part of the classical information (e.g., the employed basis for the quantum state preparation and measurement), use this information to select a subset of the detection events (of size $N_s \leq N_d$) and employ some of the corresponding data (e.g., the measurement outcomes when Alice and Bob employ matching bases) to extract the raw keys, $k_\upA$ and $k_\upB$. These keys are bit strings of length $\ell_\text{raw} = N_s b_q$, where $b_q$ is the number of sifted bits per detected quantum signal. The key mapping function might involve further random choices (e.g., applying a random permutation to the order of the sifted key bits) or compress the data (e.g., mapping a continuous value to a discrete value in constinuous-variable protocols).

\item \textit{Parameter estimation:} Alice and Bob may use part of the data to estimate the quality of the quantum signal (e.g., the QBER in one basis). These estimations may be employed to tune the protocol parameters in the next steps.

\item \textit{Information reconciliation (a.k.a., error correction):} The raw keys $k_\upA$ and $k_\upB$ extracted by Alice and Bob typically differ on a fraction of the bits. Some information is exchanged on the classical channel to run an error correction protocol and yields keys $k_\upA'$ and $k_\upB'$ which are equal on Alice' and Bob's side with high probability.\footnotemark 

\item \textit{Error verification (a.k.a., hashing):} Alice and Bob apply a hash function to the respective keys ($k_\upA'$ and $k_\upB'$) and keep them only if the resulting hashes are equal. The hash function parameters are chosen so that the remaining keys are equal, except with $\varepsilon_\text{corr}$ probability of hash collision.

\item \textit{Privacy amplification:} An upper bound to the amount of information that Eve may posses is derived. A suitable 2-universal hash function is applied to the key held by Alice and Bob yeilding a bit string of length $\ell \leq \ell_\text{raw}$. According to the quantum left-over hash lemma\cite{tomamichel2011leftover}, the result will appear as a uniformly random bit string from Eve's perspective, except for $\varepsilon_\text{sec}$ failure probability, in which case the entire key may be leaked.
\end{enumerate}}
\end{tcolorbox}

\footnotetext{After information reconciliation the probability $\delta$ that the keys differ, $\delta = \Pr(k_\upA' \neq k_\upB')$, is usually in the range $10^{-2} - 10^{-6}$ and depends on the QBER and on the adopted error correction scheme. Hashing is then employed to discard unequal keys. Thus, a high value of $\delta$ decreases the protocol success probability, but does not compromise the protocol security.}

The protocol may abort in each of the post-processing steps 2,3,4,5 if certain conditions are not met (e.g., if the sifted key is too short, if the QBER is too high, or if either of the error correction and verification step fails). In case of abort, the currently processed block of information is discarded by both Alice and Bob, otherwise a secure key of length $\ell$ is obtained. For a fixed quality of the quantum channel (specified, e.g., by the end-to-end transmission and the QBER) the expected length of the secure key depends on the total number of exchanged quantum signals and the employed security parameter, $\avg{\ell} = \textup{SKL}(N,\varepsilon_\text{sec},\varepsilon_\text{corr})$. More signals results in longer keys, while more stringent security and correctness parameters result in shorter secure keys.

\section{Security considerations in satellite QKD systems}
\label{sec:security}

In this section we introduce some of the factors determining the security of a satellite QKD system from different perspectives, including the QKD protocol selection, its physical implementation, and the communication network architecture.

\subsection{Assumptions on the eavesdropper attack model}

The baseline assumptions on the eavesdropper attack model in QKD are as follows.
\begin{enumerate}
\item Eve has arbitrarily large computing power, thus the security shall not hinge upon computational hardness assumptions.
\item The classical and quantum channels are both under Eve's control (but the classical messages are authenticated). She can perform arbitrary operations on the QKD signals, limited only by the laws of quantum mechanics, and can store an arbitrary amount of quantum and classical information.
\item The QKD devices owned by Alice and Bob are well-characterised and only a small deviation from their mathematical description is allowed. Furthermore, the QKD devices are within Alice and Bob's secure perimeters and thus assumed to be inaccessible to Eve.
\end{enumerate}
Each of these three conditions may be modified, further strengthening or relaxing the QKD security. 

The first condition may be relaxed, e.g.\ allowing PQC signatures for authentication\cite{wang2021experimental}. The second may be strengthened, e.g.\ basing the security only on no-signaling\cite{barrett2005no}; modified by introducing relativistic constraints\cite{sandfuchs2023security}; or weakened by assuming that the Eve has only partial access to the quantum channel\cite{ghalaii2023satellite}. The third one may be strengthened, e.g.\ by using device-independent (DI) approaches\cite{primaatmaja2023security}; or weakened, by introducing intermediate Trusted Node (TN) for relaying the key\cite{salvail2010security}. 

Some of these modifications could play an important role in SatQKD. For instance, relativistic principles might be exploited in the security proof: knowing precisely the positions of the satellite and of the Optical Ground Station (OGS) one can put stringent upper bounds to the maximum time delay that Eve could have induced in the flying photons, $\delta t_\text{Eve}$; if the photons are received at a rate lower than $1/\delta t_\text{Eve}$ Eve can interact with at most one photon at a time, which then allows one to exploit novel entropy accumulation theorems to prove security\cite{metger2023security}. Another possibility is to exploit the fact that in SatQKD it is very difficult for Eve to collect all the transmitted photons, since she would need to completely block the line of sight between the transmitter and receiver to do so. Thus she may have access to a fraction of the signals (a so-called \textit{bypass channel}) and, under realistic assumptions, this can lead to significantly improved system performance\cite{ghalaii2023satellite}.

\subsection{Maturity of security proofs for SatQKD}

Historically, the first QKD security proofs were given against restricted classes of attacks for which it is easier to derive bounds to the information accessible to Eve.
One can identify \textit{individual}, \textit{collective}, and \textit{coherent} attacks\cite{pirandola2020advances}. 
In individual attacks, Eve acts independently on the quantum signals sent in each communication round; in \textit{collective attacks}, Eve first acts individually on each quantum signal, then performs a collective measurement on her quantum system that stores the quantum information she has collected; and in \textit{coherent attacks}, the most general attack in the first assumptions listed above, Eve can store all the quantum signals simultaneously and perform arbitrary quantum information processing on these.
Albeit only individual attacks are feasible with currently existing technology, in the present work consider only protocols for which a security proof against of coherent attacks is available. The goal is to select protocols for which the security proofs are more mature and allowing a comparison of the protocols on an equal ground.

A complete security proof should also account for all the (known) discrepancies between the mathematical description of the QKD protocol and its physical implementation. Some distinctive technological features have been directly incorporated in the protocol description, such as the use of Weak Coherent Pulses (WCP) rather than Single-Photon Sources (SPS) in discrete-variable QKD (solved with the introduction of decoy-states\cite{lo2005decoy}), or the use of threshold detectors rather than photon-number resolving detectors (requiring the definition of squashing models\cite{gittsovich2014squashing}). Other effects have instead to be modelled as (small) deviations from the specified behaviour and their impact on the achievable key rate evaluated. These include, e.g.: spectral, spatial, or temporal distinguishability of the transmitted quantum states; patterning effects (i.e., spurious correlations between consecutive quantum signals) due to the finite bandwidth of the control electronics; the mismatch in detector efficiency; and many more. Despite years of theoretical investigations, no security proof currently cover all these aspects simultaneously\cite{gottesman2004security, makarov2023preparing}. The numerical security framework based on semi-definite programming\cite{coles2016numerical, winick2018reliable, george2021numerical} appears to be the one closest to being able to close all the gaps in the security proofs simultaneously. Lacking a systematic way to address these issues, in this work we will only employ analytical expressions for the key generation rate and only asses the impact of photon losses and QBER.

\subsection{Device-independent approaches}

Secret keys can be generated either via a direct link between Alice's and Bob's QKD devices or employing an intermediate node, which may perform either entanglement distribution or entangling measurements. Depending on the quantum link configuration one can thus identify three families of QKD protocols: Prepare-and-Measure (PM), Entanglement-Based (EB) and Measurement-Device-Independent (MDI).

In PM protocols a QKD transmitter directly sends quantum signals to a QKD receiver. Both devices have to be well characterized and to be trusted by the end-users. 
In EB protocols a intermediate node hosts an entanglement source and is connected via quantum channels to Alice and Bob: entangled quantum states are distributed to Alice and Bob, who both employ a QKD receiver to measure them. MDI protocols can be regarded as the dual of EB protocols: Alice and Bob use QKD transmitters and they simultaneously send quantum states towards the intermediate node which performs entangling measurements on them.\footnote{In PM protocols the QKD transmitter and receiver are typically called Alice and Bob modules, respectively. This terminology is not adopted in EB and MDI protocols, as in these cases the equipment at the end-points is functionally the same.}

EB-QKD and MDI-QKD only require an intermediate UnTrusted Node (UTN). Some famous EB protocols are the one introduced by Ekert in 1991 (E91)\cite{ekert1991quantum} and by Bennett, Brassard, and Mermin in 1992 (BBM92)\cite{bennett1992quantum}, while the first MDI protocol was introduced by Lo, Curty, and Qi in 2011\cite{lo2012measurement}. Trust in the entanglement source and in the measurement device, respectively, are not required: the security of the QKD exchange is established by the protocol itself during the data post-processing. In fact, the security proofs are set in the worst-case scenario where the intermediate node is in Eve's hand, who may perform arbitrary state preparations and measurements. Similarly as done in a PM protocol, Alice and Bob disclose the value of part of the locally generated variables (i.e., the measurement bases and outcomes for EB-QKD and the settings used for state preparation in MDI-QKD) and perform parameter estimation. If the QBER is sufficiently low it is possible to upper bound the information that Eve may posses and ultimately extract a secure key.

Finally, in fully device-independent (DI) approaches all quantum devices are treated as black boxes that are not trusted by the end-users. Effectively, the end-user assume that the devices have been manufactured by the Eve, the untrusted party. Even in this very pessimistic scenario one can run a secure QKD protocol provided that: (1) Alice and Bob can stochastically set the inputs of their QKD devices and (2) they can isolate the QKD devices from the external environment, so that cannot illicitly leak any information to Eve\cite{primaatmaja2023security}. Recently, the first experimental demonstrations of DI-QKD have been carried out\cite{zapatero2023advances}. Nonetheless, the technical challenges of implementing DI-QKD are formidable and therefore this class of protocols will be left out of our analyses.

\subsection{Hardware security and side channels}

QKD transmitter and receiver modules are prone to having side-channel open (either classical or quantum), which may inadvertently leak some or all the information to Eve. Side-channel attacks can be active or passive, i.e., may require or not the active use of probe signals by Eve to acquire the information. Unless device-independent approaches are employed, it is paramount to achieving security that the devices are very thoroughly analysed and characterised, to the point where no room for unknowns is left over. A thorough investigation of all the known vulnerabilities has recently been made by the German Federal Office for Information Security (Bundesamtes für Sicherheit in der Informationstechnik, BSI), showcasing the amount effort that is required to achieve security in practice\cite{BSI}. 

SatQKD inherits some of its strengths, compared to ground-based QKD, from the use of satellites as protected hardware. Once in space, it requires a huge effort to get physical access to the hardware for intentional manipulation, though not impossible in principle; for insatnce, see the co-orbital spacecraft event of a payload released by the Russian satellite COSMOS 2570 recently detected by LeoLabs\cite{spoofing}. The stringent  procurement process from traceable sources, as well as extensive testing, provide a higher level of trust in the security of the critical security hardware components, compared to non-space-qualified technology. However, that doesn't mean that the absence of back-doors or otherwise malicious functions can be guaranteed.

Threat actors may use \textit{classical} side-channel attacks to gather information or influence the execution of a (sub-)system by measuring or exploiting indirect effects. For instance, an active side-channel attack could consist in inducing a fault in an intermediate variable (i.e., the result of an internal computation) of a cipher by applying an external stimulation on the hardware during runtime, such as a voltage or clock glitch. As a result of fault injection, specific features appear in the state of sensitive variables under attack\cite{willbold2023space}. But also purely passive attacks could be used to extract information through side-channel attacks. For instance, an electrical component, such as a step-motor steering a mirror for the optical path, could be monitored to identify which of the phase elements are in the optical path at any time. Thus it is of upmost importance to make sure in the system design that no electrical, thermal, magnetic or any other kind of signal is emitted or can be measured by any other means which would reveal the internal status of the quantum and non-quantum elements onboard\cite{baliuka2023deep}.

As a matter of illustration we also present two well-known \textit{quantum} side-channel attacks, one on the transmitter side and one on the receiver side. In the Trojan-horse attack Eve injects strong laser pulses into the QKD transmitter in order to ascertain the state of the modulators by how they affect the back-reflected light\cite{gisin2006trojan}. Counter-measures are possible (employing optical isolators and monitoring the incoming light) and should be put in place. In the double-click attacks, Eve sends strong pulses with a given polarisation towards the QKD receiver; this allows her to gain control of the receiver, if the events where both detectors in a given basis simultaneously click are discarded\cite{lutkenhaus1999estimates}. This and some other detector attacks can be solved by post-processing of the signal. Specifically, if a \textit{squashing model} exists, then the behaviour of the \textit{real} detector can be proven to be mathematically equivalent to that of in \textit{ideal} detector which is not vulnerable to these attacks\cite{gittsovich2014squashing}. For further information on possible active and passive attacks we refer to the BSI document\cite{BSI}.

\subsection{Satellites as trusted nodes}

When it is impossible to have a direct QKD link between the end-users (e.g., due to exceedingly high transmission losses), several consecutive short-range QKD links may be established, and a QKD key can be relayed through these links. Since at the end-points of each link the QKD key is (potentially) available in unencrypted form, all the intermediate nodes should be trusted by the end-users not to leak secret information. Under these conditions, TNs can be employed to securely extend the effective communication range of QKD links\cite{salvail2010security}. 

We argue that satellites employed as TNs offer excellent practical security, compared to terrestrial TN networks. One advantage is that it is very hard, with current technology, to access a satellite without being detected, and only few attacks of this kind have been attempted until now\cite{spoofing}. Furthermore, as will be discussed later, a single satellite employed as TN could link any pair of end-users located anywhere on Earth, while with fibre-based links dozens or hundreds of intermediate nodes may be required to bridge intercontinental distances; therefore, the surface of attack is significantly reduced. Finally, TN on the ground face the conundrum that they should be inaccessible to potential eavesdroppers, but be accessible to service personnel for upgrade and maintenance operations.

\section{Current status of satellite QKD}
\label{sec:status}

\subsection{Generic optical link architecture}

\begin{figure*}[t!]
\centerline{\includegraphics[width=0.95\textwidth]{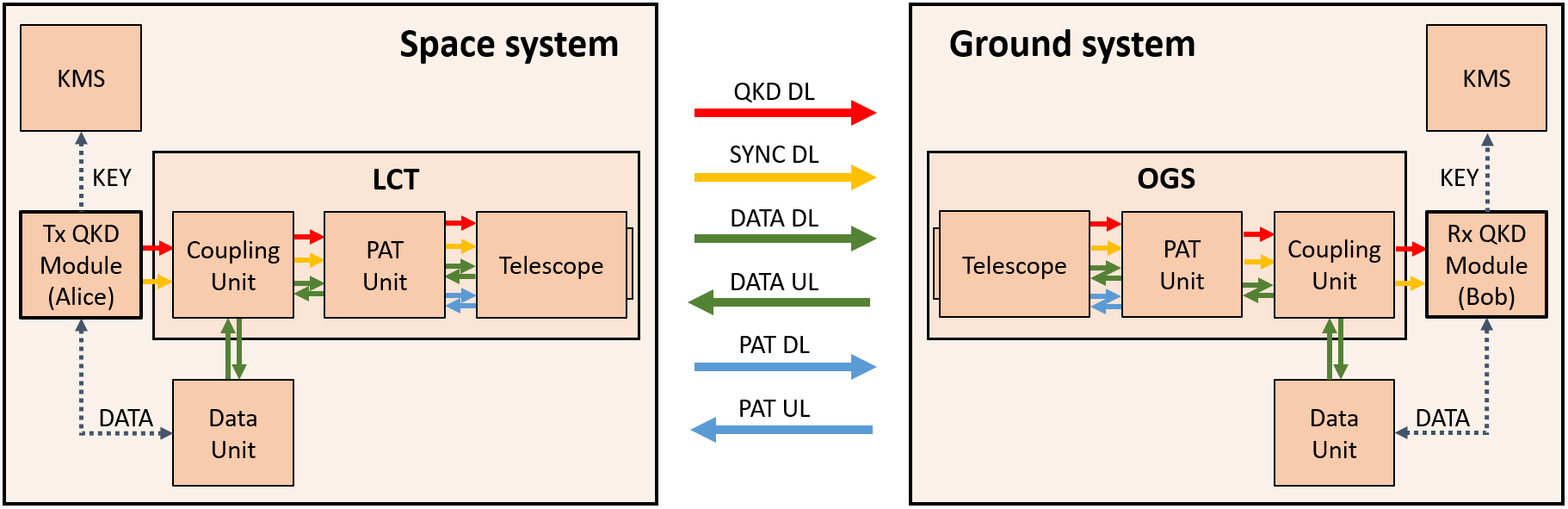}}
\caption{Optical link architecture of a generic PM SatQKD system. The quantum channel downlink direction is chosen for simpler visualization; in the case of a quantum channel uplink the red and orange arrows are reversed. Solid arrows denote optical signals, dashed arrows electric signals. 
\label{fig:linkArch}
}
\end{figure*}

Aside from the quantum channel, the classical support functionalities for SatQKD also require FSO links. These include classical data communication, laser beaconing, signal time synchronisation and polarisation (or phase) alignment. In particular, a high-throughput classical communication channel is needed for QKD post-processing and should be realised via FSO links.\footnote{RF links may have a supporting role, e.g.\ for telemetry and telecommand.} The system requirements of classical support functionalities are comparatively less demanding than those of the quantum channel (which employs signals that are several orders of magnitude weaker), but they nonetheless affect the system architecture.

The optical link architecture of a satellite QKD systems consists of the optical channels of the satellite QKD system and the respective sources and sinks. It is important to carefully select the optical channels because these have significant impact on the overall system design and performance\cite{kaushal2016optical}. In particular, large wavelength separation between the quantum link and the classical channels is advised, to allow minimisation of QBER to channel cross-talk via spectral filtering. Furthermore, some of the classical functionalities may be combined together in order to reduce system complexity.

The sketch in Figure~\ref{fig:linkArch} shows a generic optical link architecture of a prepare-and-measure SatQKD system in DownLink (DL). The Uplink (UL) view would be similar and only change the direction of QKD link and Sync link, while the links employed for Pointing, Acquisition and Tracking (PAT) and for classical data communication (DATA) remain the same. Together, the Alice and Bob module in the space system and the ground system comprise the actual QKD system. The final key is stored and managed in the Key Management System (KMS). The data units comprise the classical data sources and sinks, together with the modules transducing electrical signal to and from the optical domain. The space system coupling unit combines the optical signals from the Alice module and the data unit with output to the Laser Communication Terminal (LCT). The LCT performs beam forming and steering and points towards the OGS which captures the beam and guides it to the respective coupling unit. This one splits the different optical channels and outputs to the Bob module and the data unit.

\subsection{Technological and operational challenges of satellite QKD}

Distribution of quantum-generated keys via satellite links entails a series of technological and operational challenges compared to QKD via optical fibre links; none of these is fundamentally insurmountable, but they do result in a significant increase of the overall system complexity. However, SatQKD is the only method that can enable, with currently existing technology, QKD connections for end-users at continental and intercontinental distances, thus justifying the effort of developing and deploying SatQKD systems.

From the technological standpoint, several subsystems have to be added to enable the establishment of a dynamical FSO link\cite{sidhu2021advances}. In particular, a PAT system has to be included to establish and maintain the quantum link for as long as needed, while compensating platform vibrations and the change in link geometry during the satellite pass. This usually calls for the use of dedicated laser beacon systems for each FSO link, each mounted on one of the two end-points, allowing the terminals to track the position of one another. Furthermore, the same optical terminal must be employed for receiving both the quantum signal and the classical support signals, including functionalities such as classical data communication, beaconing and time synchronisation; therefore, it must be possible to separate them only by means of spectral or temporal distinguishability, since the beams will be overlapping at the external aperture of the receiver and thus spatial separation is not possible. The QKD payload itself has to fulfill a series of extra requirements for it to be mounted on a satellite. These include compatibility with the Size, Weight and Power (SWaP) requirements of the satellite bus which, especially for CubeSat platforms, can be rather stringent. It must be possible to interface the QKD module to the FSO communication system (for classical communications), either directly with free-space components (i.e., with a coudé path) or by providing an optical fibre interface. Furthermore, it must be able to operate in the harsh space environment, subject to vacuum conditions, large thermal excursions, high radiation exposure and magnetic field fluctuations. This is further complicated by the fact that it is essentially impossible to perform maintenance, reparations and calibrations once the equipment is flown into space. Last but not least, the QKD payload must be robust enough to survive the rocket launch, which involves strong accelerations and intense shaking.

From the standpoint of quantum link operations, SatQKD faces the following challenges\cite{sidhu2022finite}. First, high transmission losses are to be expected, mainly due to beam divergence and the large communication distances resulting in a beam spot on the ground that is typically much larger than the telescope collection area. Second, the link duration between a satellite in Low-Earth Orbit (LEO) and an OGS is limited to those few minutes per day in which the satellite is passing above the OGS's horizon and a direct link can be established. Third, the QKD exchange has to succeed in the presence of large channel fluctuations, including variable free-space loss due to changing link distance, variable atmospheric absorption due to changing elevation angle (slow predictable variations) and losses due to pointing jitter and turbulence-induced optical power scintillation (fast unpredictable variations). Fourth, the presence of cloud coverage and fog will result in stochastic service unavailability. Finally, FSO links are impacted by the presence of background light and appropriate filtering of the quantum signal in the spectral, spatial and temporal domain should be applied. Operations during the day are very challenging due to the presence of sunlight. While this may be possible using purposefully designed systems (requiring excellent signal filtering and high antenna gains), in this paper we will only consider night-time operations. By night, the background light contributions stem from from the Moon, the stars and human activities and are orders of magnitude lower than during daylight. Furthermore, classical FSO support system cannot be spatially separted from the quantum channels. The cross-talk with the classical channel can then result in an increase of the QBER.

\subsection{Status of research and development in satellite QKD}

In this section the status of the global research and development efforts is described. This is done by a selection of important scientific articles and technology demonstrator missions. The selection does not claim to cover all relevant articles and missions but shall show maturity of research, development and demonstration of the technology. Table~\ref{tab:reviews1} contains some of the topical reviews and fundamental papers on SatQKD which are important to understand the current status of research. Table~\ref{tab:reviews2} lists  planned and flying QKD satellites. Further information can be found in the reviews and tutorial type papers given by Bedington \cite{bedington2017progress}, Sidhu \cite{sidhu2021advances}, Scriminich \cite{scriminich2022optimal} and Lu \cite{lu2022micius}.

\begin{table}[t!]
\begin{tabular}{|p{0.12\textwidth}|p{0.03\textwidth}|p{0.15\textwidth}|p{0.60\textwidth}|}
\hline
\textbf{Main topic} & \textbf{\!Ref.} & \textbf{First author, year} & \textbf{Short description} \\
\hline\hline

\multirow{4}{0.15\textwidth}{\\[15mm]General\\SatQKD\\reviews}
 & \vspace{1pt}\cite{bedington2017progress}
 & \vspace{-2pt}\href{https://doi.org/10.1038/s41534-017-0031-5}{Bedington, 2017} 
 & Introduction to SatQKD with TN; short overview of QKD technologies for single photon sources, link tracking and photon detection.
 \\ \cline{2-4}
 & \vspace{1pt}\cite{sidhu2021advances}
 & \vspace{-2pt}\href{https://doi.org/10.1049/qtc2.12015}{Sidhu, 2021} 
 & Extensive review of current and planned QKD CubeSat missions; history of FSO QKD; concepts of satellite-based entanglement distribution networks.
 \\ \cline{2-4}
 & \vspace{1pt}\cite{scriminich2022optimal} 
 & \vspace{-2pt}\href{https://doi.org/10.1088/2058-9565/ac8760}{Scriminich, 2022} 
 & Design and performance evaluation of SatQKD with system parameters optimisation; detailed atmospheric channel modelling; adaptive optics corrections at finite bandwidth; finite-key effects.
 \\  \cline{2-4}
 & \vspace{1pt}\cite{lu2022micius}
 & \vspace{-2pt}\href{https://doi.org/10.1103/RevModPhys.94.035001}{Lu, 2022} 
 & Summary of the quantum communication experiments performed by the Micius satellite; included is also a general review of SatQKD; . \\ 
\hline

\multirow{2}{0.15\textwidth}{\\[2mm]Atmospheric\\channel}
 & \vspace{1pt}\cite{bonato2009feasibility}
 & \vspace{-2pt}\href{https://doi.org/10.1088/1367-2630/11/4/045017}{Bonato, 2009} 
 & Early feasibility study of SatQKD, both in uplink and downlink; theoretical model of background light; polarisation control.\\ \cline{2-4}
 & \vspace{1pt}\cite{vasylyev2019satellite}
 & \vspace{-2pt}\href{https://doi.org/10.1103/PhysRevA.99.053830}{Vasylyev, 2019} 
 & Study of atmospheric effects in SatQKD; link range including beam refraction; atmospheric attenuation; aperture-averaged scintillation. \\
\hline

\multirow{2}{0.15\textwidth}{\\[2mm]Network\\optimisation}
 & \vspace{1pt}\cite{polnik2020scheduling}
 & \vspace{-2pt}\href{https://doi.org/10.1140/epjqt/s40507-020-0079-6}{Polnik, 2020} 
 & Scheduling of SatQKD downlink to serve multiple OGS; formulated as a constrained, weighted key maximisation problem. \\ \cline{2-4}
 & \vspace{1pt}\cite{erhard2021choose}
 & \vspace{-2pt}\href{https://doi.org/10.1117/12.2599218}{Erhard, 2021} 
 & Comparison of three SatQKD scenarios (BB84, BBM92, TF-QKD); summary of the technological and encoding choices in SatQKD.\\ 
\hline 

\multirow{3}{0.15\textwidth}{\\[4mm]Finite-size\\effects}
 & \vspace{1pt}\cite{lim2021security}
 & \vspace{-2pt}\href{https://doi.org/10.1103/PhysRevLett.126.100501}{Lim, 2020} 
 & Security proof with tight key rate analysis; minimisation of finite-size overhead for QKD with small block lengths, as typical in SatQKD. \\ \cline{2-4}
 & \vspace{1pt}\cite{sidhu2022finite}
 & \vspace{-2pt}\href{https://doi.org/10.1038/s41534-022-00525-3}{Sidhu, 2022} 
 & Assessment and optimisation of finite key effects in SatQKD: averaging of QBER over a satellite pass is required, signals sent at low link elevations should be discarded.\\ 
\hline

\multirow{4}{0.15\textwidth}{\\[15mm]Papers on\\CubeSat\\missions}
 & \vspace{1pt}\cite{bedington2016nanosatellite}
 & \vspace{-2pt}\href{https://doi.org/10.1140/epjqt/s40507-016-0051-7}{Bedington, 2016} 
 & Early study of CubeSat platforms to enable SatQKD missions; space-compatible entanglement sources. \\ \cline{2-4}
 & \vspace{1pt}\cite{oi2017cubesat}
 & \vspace{-2pt}\href{https://doi.org/10.1140/epjqt/s40507-017-0060-1}{Oi, 2017} 
 & CubeSat for quantum communications; preliminary payload design; concepts of operations; optics and fine pointing assembly.\\ \cline{2-4}
 & \vspace{1pt}\cite{kerstel2018nanobob}
 & \vspace{-2pt}\href{https://doi.org/10.1117/12.2535988}{Kerstel, 2019} 
 & CubeSat for QKD uplink; feasibility study; preliminary payload design; pointing, acquisition and tracking.\\  \cline{2-4}
 & \vspace{1pt}\cite{islam2022finite}
 & \vspace{-2pt}\href{https://doi.org/10.48550/arXiv.2204.12509}{Islam, 2022} 
 & Comparison of three proposed CubeSat QKD donwnlink missions; improvements in finite-key effects analysis allow producing secure keys even for high transmission losses.\\ \cline{2-4}
 & \vspace{1pt}\cite{zhang2023end}
 & \vspace{-2pt}\href{https://doi.org/10.48550/arXiv.2312.02002}{Zhang, 2023} 
 &  System level prototype of SatQKD terminal for low-cost nano-satellites; discussion of background noise, gate width and mean photon number on QBER and SKR.\\
\hline
\end{tabular}

\caption{List of recent paper about SatQKD and related topics. This is not meant to be a comprehensive list of studies, but only a sample of the current activities in the field.}
\label{tab:reviews1}
\end{table}

A variety of quantum satellite missions employing the SatQKD architecture selected in this work are already in space or on their way. Many of these missions support the BB84 protocol which underlines the significantly and maturity of the protocol. The satellites in space hosting a respective payload are MICIUS, Tiangong-2, and Jinan 1. Key generation could be successfully demonstrated with MICIUS and Tiangong-2, whereas for Jinan~1 no experiments are reported till now.

Some further missions, planned for launch, include QUBE-II, Eagle-1, IRIS$^2$, and QEYSSat. These missions showcase the international efforts and interest in SatQKD. 

QUBE-II is a CubeSat hosting two different QKD transmitters for experimental demonstration of LEO downlink BB84, with launch planned for 2025\cite{Hutterer2022}. Eagle-1 includes a QKD system and goes beyond pure experimental demonstration by implementing an operational key service with launch planned till 2025\cite{SES2022}. IRIS$^2$ is the Secure Space Connectivity initiative by the European Union and will integrate the European Quantum Communication Infrastructure with its space component SAGA (Security And cryptoGrAphic Mission). A SAGA First Generation mission will contain one LEO satellite with a PM-QKD system with launch planned till 2027 for in-orbit validation of, amongst other, QKD performance profiling, system verification and service validation. Key parameters of the QKD protocol implementation are polarisation encoding, prepare-and-measure type, discrete variable, quantum channel DL, wavelength in C-band \cite{Lindman2023}. A future generation of the system will be integrated into the EU Secure Satellite Constellation IRIS$^2$ (Infrastructure for Resilience, Interconnectivity and Security by Satellite) \cite{Steiner2023}. QEYSSat is a research quantum satellite that enables uplink quantum communication \cite{Scott_2020}. Table~\ref{tab:reviews2} lists the QKD satellites that are already in space and a selection of satellites to be launched within the next few years.

\begin{table}[t!]
\begin{tabular}{|p{0.14\textwidth}|p{0.03\textwidth}|p{0.13\textwidth}|p{0.60\textwidth}|}
\hline
\textbf{Status} & \textbf{\!Ref.} & \textbf{Satellite} & \textbf{Short description} \\
\hline\hline

\multirow{4}{0.15\textwidth}{\\[12mm] QKD satellites\\in space}
 & \vspace{1pt}\cite{2017Liao} 
 & \vspace{-2pt} Micius 
 & First satellite to demonstrate QKD from space; quantum sources support prepare-and-measure QKD, entanglement based QKD and further quantum communications experiments. 
 \\ \cline{2-4}
 & \vspace{1pt}\cite{Liao2017} 
 & \vspace{-2pt} Tiangong-2 
 & Chinese space station which hosted a compact QKD payload and performed downlink to several ground stations.
 \\ \cline{2-4}
 & \vspace{1pt}\cite{Jones2023} 
 & \vspace{-2pt} Jinan 1 
 & Dedicated QKD satellite with reduces size with respect to the Micius satellite and first test QKD satellite for the LEO network. 
 \\   
\hline

\multirow{4}{0.15\textwidth}{\\[20mm]QKD satellites\\planned}
 & \vspace{1pt}\cite{Hutterer2022}
 & \vspace{-2pt} QUBE-II
 & CubeSat with two different PM-QKD systems, implemented with wavelengths around \SI{850}{\nano\meter} and \SI{1550}{\nano\meter}. \\ \cline{2-4}
 & \vspace{1pt}\cite{SES2022}
 & \vspace{-2pt} Eagle-1 
 & SatQKD system to be developed as partnership among the European Commission, the European Space Agency (ESA) and various European space companies.\\  \cline{2-4}
 & \vspace{1pt}\cite{Lindman2023}
 & \vspace{-2pt} SAGA 1G
 & SAGA 1G is the first generation of the space segment of the European Quantum Communication Infrastructure (EuroQCI).\\  \cline{2-4}
 & \vspace{1pt}\cite{Steiner2023} 
 & \vspace{-2pt} IRIS$^2$ 
 & IRIS$^2$ (Infrastructure for Resilience, Interconnectivity and Security by Satellite) is the EU Secure Satellite Constellation which will integrate the EuroQCI.\\ \cline{2-4}
 & \vspace{1pt}\cite{Scott_2020}
 & \vspace{-2pt} QEYSSat
 & QEYSSat will carry a QKD science payload to test the BB84 and BBM92 protocols. In contrast to other missions, the quantum communication is in uplink.  
 \\ 
\hline 
\end{tabular}

\caption{List of QKD satellites that are currently in space or planned. This is not meant to be a comprehensive list of satellite QKD missions, but only a sample of the current activities in the field.}
\label{tab:reviews2}
\end{table}

\section{Definition of the reference satellite QKD architecture}
\label{sec:architecture}

It is possible to categorise SatQKD implementations according to several different criteria and most features can be independently selected and combined, resulting in exponentially many potential architectures. The interplay between the selected features can be complex, but we consider them mostly in isolation to keep the exposition simple. We determine the SatQKD architecture that we deem as most promising for near-term implementations, thus restricting the space of possibilities to be further analysed. In this section we thus motivate the following architectural choices.

\begin{tcolorbox}{
\noindent\textbf{Architecture definition}
\begin{enumerate}
    \item Satellites in Low-Earth Orbit (\textbf{LEO}) are preferred since they yield higher key generation rates rather than those in Medium-Earth Orbit (MEO) or GEOstationary orbit (GEO).
    \item The use of prepare-and-measure (\textbf{PM}) is preferred over QKD protocols where the satellite acts as a (not necessarily trusted) third party, such as Entanglement-Based (EB) and Measurement-Device-Independent (MDI) protocols.
    \item Consequently, the satellites have to be employed as Trusted Nodes (\textbf{TN}), since the UnTrusted Node (UTN) functionality does not exist in PM configurations.
    \item For the link direction of the quantum signal, DownLink (\textbf{DL}) is more favourable than UpLink (UL), due to the higher signal transmission and lower system complexity.
    \item Among the classes of  QKD protocols, those based on Discrete Variables (\textbf{DV}) have superior performance in typical satellite links compared to those based on Continuous Variables (CV).
    \item For the transmitter technology, we restrict our analysis to the use of Weak Coherent Pulses (\textbf{WCP}) over Single-Photon Sources (SPS), since the former is based on very mature laser technology.
    \item In the class of decoy-state PM-DV-QKD protocols \textbf{BB84} appears to be the best candidate due to its low implementation simplicity, high key generation rate, and comparatively few gaps in its implementation security.
\end{enumerate}}
\end{tcolorbox}

This architecture has emerged in the aerospace and quantum communication community as the most promising for the near-term implementation of SatQKD. Naturally, many other possible architectures are of technological interest and worth being investigated. The SatQKD field is still in an early development phase and other architectures may prove advantageous in the future for certain application scenarios.

\subsection{Selection of the satellite orbit height (LEO, MEO, GEO)}
\label{sec:LEO}

For simplicity we only consider circular orbits, which are fully specified by three parameters: $R_\sat$ the orbit radius ($R_\sat = R_\oplus + h$, where $R_\oplus = \SI{6378.137}{\kilo\meter}$ is \textit{Earth’s equatorial radius} and $h$ is the satellite altitude); $i$, the orbital plane inclination; and $\varphi$, the Right Ascension of the Ascending Node (RAAN). The orbits under considerations are thus LEO ($h \leq \SI{2000}{\kilo\meter}$), including Very-Low-Earth Orbits (VLEO, $h \leq \SI{400}{\kilo\meter}$), MEO ($\SI{2000}{\kilo\meter} \leq h < \SI{35786}{\kilo\meter}$) and GEO ($h = \SI{35786}{\kilo\meter}$ with inclination $i \simeq 0^\circ$). Here we discuss the contrasting factors that influence the choice of the best satellite altitude. 

\paragraph*{Challenges of low-altitude orbits}
On the one hand, employing lower orbits presents a number of challenges. The ground area covered by a satellite is $A_\text{cov} = 2\uppi R_\oplus^2 \frac{h}{R_\oplus + h} \approx 2\uppi R_\oplus h$, i.e. approximately linear in $h$ for $h \ll R_\oplus$; the same applies to the globally averaged link availability, which can be computed as $\avg{F_\text{avail.}} \triangleq A_\text{cov}/A_\oplus = \frac{1}{2}\frac{h}{R_\oplus + h}$. The total time above the horizon of a satellite pass scales as $O(\sqrt{h/R_\oplus})$ for $h \ll R_\oplus$ (but also depends on the maximum elevation angle reached), lasting only a few minutes in LEO. The required satellite and OGS pointing agility, in terms of angular velocity and angular acceleration of the pointing direction, increase linearly and quadratically in $h$, respectively\cite{wertz1999space}. Finally, applying Adaptive Optics (AO) corrections becomes more challenging: the wavefront error fluctuations due to turbulence effects become faster, proportionally to the angular velocity of the satellite as seen by the OGS. Furthermore, the atmospheric beam wandering effect and the Point-Ahead Angle (PAA), corresponding to the angular distance by which the satellite moves during the OGS-to-satellite light-speed round-trip time, have to be compensated; this is challenging for low altitude orbits, since the PAA can exceed the atmospheric isoplanatic angle and therefore the beam wandering cannot be corrected only by tracking the counter-propagating beacon signal \cite{Mata_Calvo_2017}\cite{andrews2005laser}. These considerations tend to favour higher orbits, such as MEO or GEO, the latter being notoriously advantageous for having the satellite in a fixed position relative to Earth. 

\paragraph*{Advantages of low-altitude orbits}
On the other hand, choosing lower altitude orbits has some strong advantages. Most importantly, the long-term average of the key generation rate a PM-QKD protocol satisfies $\avg{R} = O(h^{-1})$ and, thus, flying a satellite at lower $h$ on average results in a producing a larger amount of secure key. The reason is the following. 
Due to beam divergence, the beam spot area grows quadratically with the link distance distance $L$ and thus, for a fixed size of the receiver telescope, the collected power decreases quadratically in $L$, $\eta = O(L^{-2})$. 
More precisely, for a diffraction-limited beam in the far-field regime the receiver collection efficiency $\eta$ is upper-bounded as\cite{klein1974optical}
\begin{align}
    \eta \leq 
    \eta_\text{coll}^\text{max} = 
    \underbrace{~G_\Tx~}_{\frac{4\uppi A_\Tx}{\lambda^2}}
    \underbrace{~G_\Rx~}_{\frac{4\uppi A_\Rx}{\lambda^2}}
    \underbrace{\eta_\text{free~space}}_{\left(\frac{\lambda}{4\uppi L}\right)^2}
    = \frac{A_\Tx A_\Rx}{L^2 \lambda^2}
    = O(L^{-2})
\end{align}
where $G_\Tx$ ($G_\Rx$) is the transmitter (receiver) ideal antenna gain, $A_\Tx$ ($A_\Rx$) is the transmitter (receiver) aperture area and $\lambda$ is the wavelength.
The average link distance is roughly linear in the satellite height, so we may simply write $\eta = O(h^{-2})$. For most PM-QKD protocols the secure key rate $R$ of PM-QKD protocols scales linearly in $\avg{\eta}$, $R = O(\avg{\eta}) = O(h^{-2})$, provided that a satellite link is available. Since the averaged link availability increases with the height, $\avg{F_\text{avail.}} = O(h)$, the long-term averaged key generation rate is thus
\begin{align}
    \avg{R} \propto 
    \underbrace{\langle F_\text{avail.} \rangle}_{O(h)} \underbrace{R(\avg{\eta})}_{O(h^{-2})}  
    = O (h^{-1}) \;.
\end{align}

Using lower orbits also has other advantages. One may employ smaller transmitter terminals to achieve a positive key rate, which imply a larger beam divergence and thus the pointing accuracy requirements are relaxed; while for smaller receiver terminals it is easier to focus the incoming signal and less advanced AO systems are required to correct the atmospheric turbulence effects. Since QKD has to be operated with quantum signals at the single-photon level, rather large (and thus expensive) telescopes and LCTs are required to perform a QKD exchange with non-zero secure key generation rate; the cost of telescopes increases very steeply with the size\cite{stahl2019multivariable} and therefore there is a drive to make optical terminals as small as possible. 

\begin{tcolorbox}{%
\paragraph*{Conclusions}
The choice of low altitude orbits is favoured as the most practical option for SatQKD applications, having increased key rates and allowing the use of realistically-sized optical terminals. We focus on the use of LEO, but not VLEO. The latter orbits are, in fact, rather impractical as the atmospheric drag increases exponentially and it has to be contrasted with active propulsion in order to avoid the de-orbiting of the satellite on a timescale of a few weeks.}
\end{tcolorbox}

\subsection{Selection of the QKD communication configuration (PM, EB, MDI)}
\label{sec:PM}

\begin{figure*}[t]
\centerline{\includegraphics[width=\textwidth]{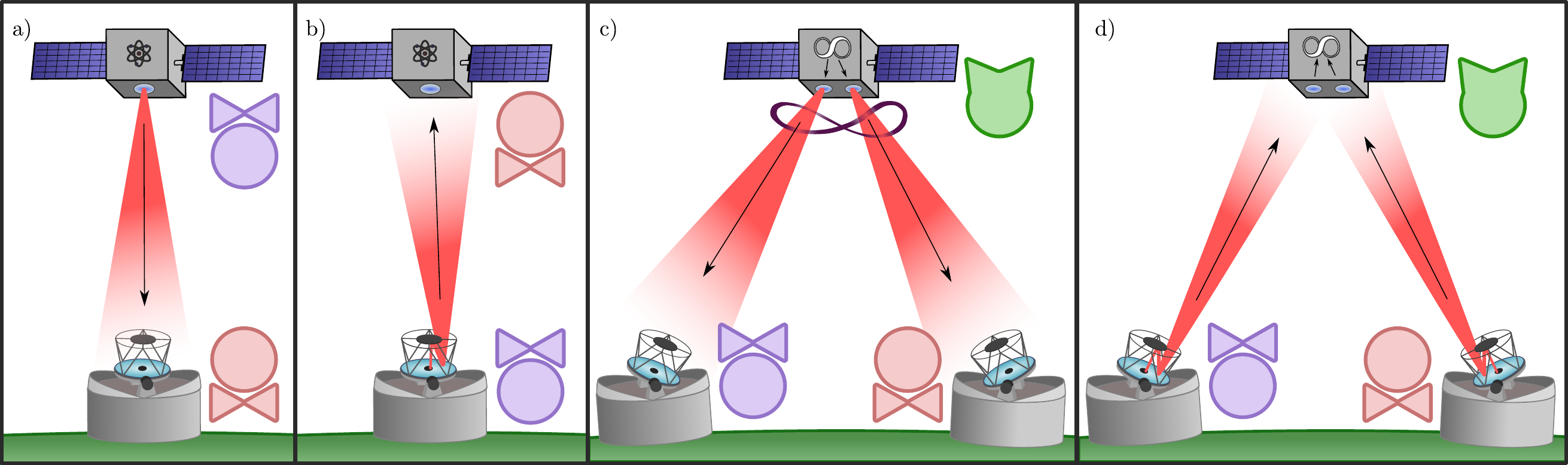}}
\caption{Possible QKD communication configurations for establishing a secure link between Alice (\raisebox{-2pt}{\includegraphics[scale=0.14]{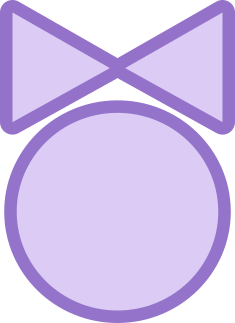}}) and Bob (\raisebox{-2pt}{\includegraphics[scale=0.14]{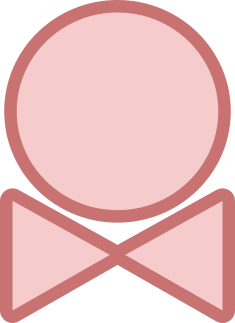}}): a) is PM-QKD protocol in DL, having the QKD transmitter in space and the QKD receiver on the ground; b) is a PM-QKD protocol in UL, having the opposite quantum link direction; c) is an EB-QKD protocol and d) MDI-QKD protocol, both having the third-party node located in space; in the last two cases the satellite does not have to be trusted and, from a security proof standpoint, may be assumed to be under Eve's control (\raisebox{-2pt}{\includegraphics[scale=0.15]{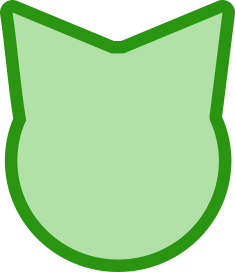}}). \label{fig1}}
\end{figure*}

The end-users of the QKD service are located in proximity of Earth's surface, while the satellites constitute third-party equipment having the role of assisting the distribution of keys between the end-users. A very natural approach is therefore to use the satellite as the central node for EB-QKD or MDI-QKD protocols. These two QKD solutions provide strong implementation security, since it only employs UTNs, i.e., no trust has to bestowed upon the satellite provider. The alternative is to use PM QKD protocols between satellite and ground, but in this case the secure key is accessible to the satellite and therefore it has to be operated as a TN. In PM protocols the communication configuration is in DL if the signals are sent from space to ground and in UL in the opposite case. See Figure~\ref{fig1} for a sketch of the four possible cases.

\paragraph*{Challenges of EB-QKD and MDI-QKD}
Despite the theoretical appeal of the EB- and MDI-QKD, since they allow foregoing the use of TNs, they result in an extreme reduction in the service that can be provided by SatQKD systems. The fundamental obstacle is that both end-users have to establish a simultaneous direct line of sight with the satellite. First and foremost, this puts a hard geometrical limit at the maximum distance at which end-users can be connected, given by\cite{wertz1999space}
\begin{align}
\label{max_dist}
    D_\text{max} = 
    2 R_\oplus \left[\arccos\!\left(\frac{R_\oplus}{ R_\oplus+h } \cos(\epsilon_\text{min})\right)-\epsilon_\text{min}\right]
\end{align}
where $h$ is the satellite height and $\epsilon_\text{min}$ is the minimum elevation angle above the horizon at which a stable link can be established; for the reference values $\epsilon_\text{min} = \SI{20}{\degree}$ and $h=\SI{574}{\kilo\meter}$ this expression evaluates to $D_\text{max} = \SI{2325}{\kilo\meter}$. Secondly, the problem of cloud blocking is further exacerbated, as it is sufficient that either one of the links is blocked to disrupt the QKD connection. Third, the complexity of the satellite optical pointing system increases substantially, since two links have to be established and maintained simultaneously. Fourth, the key rate $R$ of EB protocols and of standard MDI protocols scales as the product of each of the links end-to-end transmission efficiencies; that is, if the two links have transmissivity $\eta_\upA$ and $\eta_\upB$, respectively, the key rate scales as $R = O(\eta_\upA\eta_\upB)$. Since in typical SatQKD applications $\eta \lesssim 10^{-3}$, the key rate can be order of magnitudes lower than the one achievable when employing a single link at a time.

The very last downside could indeed be mitigated by employing innovative MDI-QKD approaches such as Twin-Field (TF) QKD\cite{lucamarini2018overcoming} and Mode-Pairing (MP) QKD\cite{zeng2022mode}, which have a secure key rate scaling as $R = O(\min\{\eta_\upA, \eta_\upB\})$, surpassing the repeaterless quantum communication bound of Pirandola, Laurenza, Ottaviani, and Banchi (PLOB bound)\cite{pirandola2017fundamental} and achieving a key rate similar to the one achievable using one repeater node. However, these novel MDI protocols are more technologically complex to realise, since they require signal phase stability. This is not required by more standard QKD protocols and is very challenging to realise in SatQKD, where the link distance dynamically changes over time.

\begin{tcolorbox}{%
\paragraph*{Conclusions}
We identify the use of PM protocols as the clearly preferred choice for near-term practical application in SatQKD. Schemes that use the satellite as the single intermediate node in EB- or MDI-QKD protocols are here rejected, since the maximum reachable communication distance by is ultimately limited by Eq.~\eqref{max_dist}. }
\end{tcolorbox}

\subsection{Satellites as trusted nodes and as untrusted nodes (TN, UTN)}
\label{sec:TN}

\begin{figure*}[t]
\centerline{\includegraphics[height=150pt]{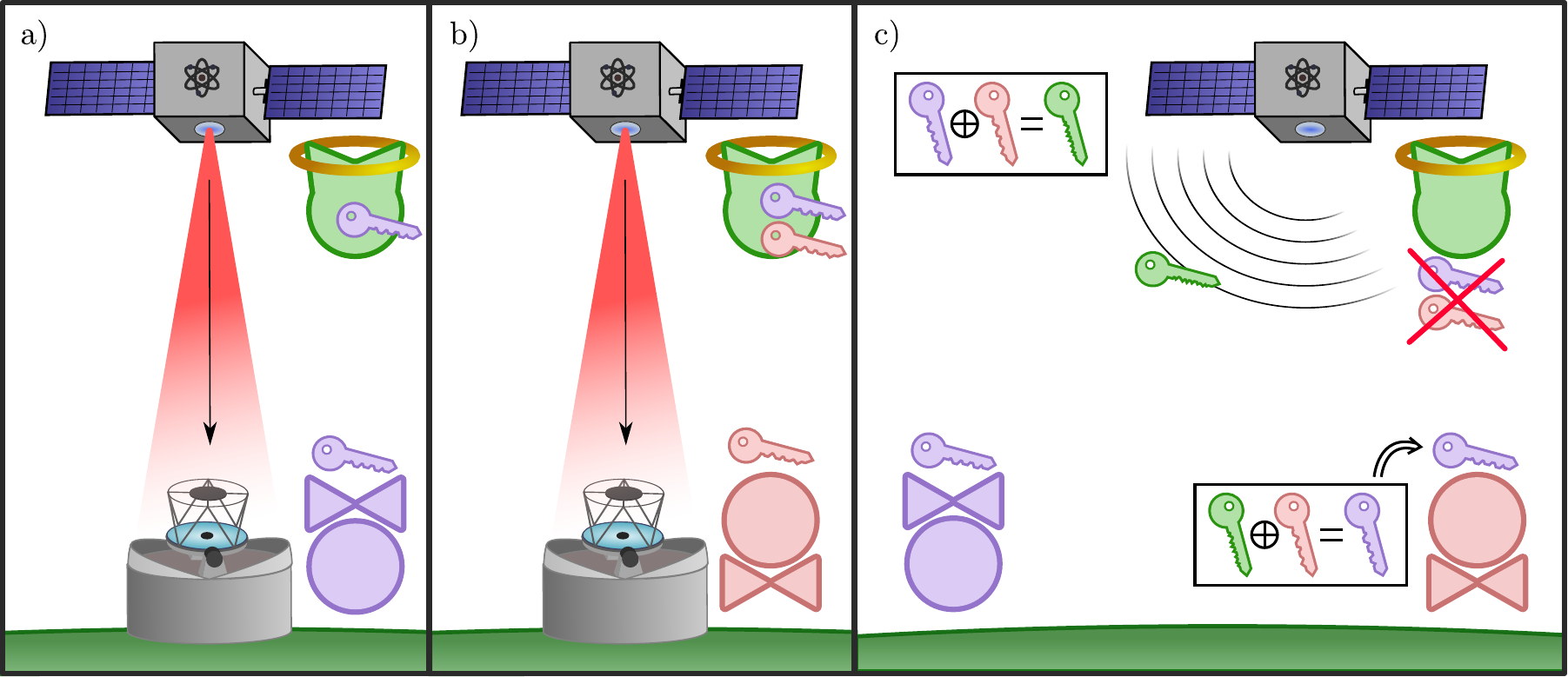}}
\caption{Operation of a QKD satellite as a trusted node; a) when the quantum link is available, a secure key $k_\upA$ is generated and shared with Alice; b) some time later, the orbit brings the satellite over Bob's location,  where a second independent key $k_\upB$ is generated and shared with him; c) at any later time, the two keys can be combined and $k_{AB} = k_\upA \oplus k_\upB$ is broadcast, so that Alice's key can securely transferred to Bob. Since $k_\upA$ and $k_\upB$ are accessible to the satellite, these have to be operated by a trusted third party (\raisebox{-2pt}{\includegraphics[scale=0.15]{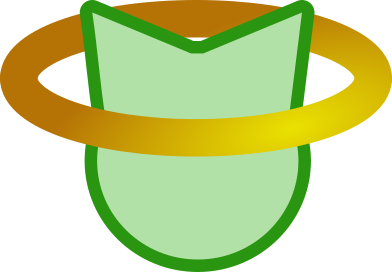}}). \label{fig2}}
\end{figure*}

We now discuss the use of TN with PM-QKD protocols in SatQKD. In the simplest form of TN-SatQKD two independently generated keys can be combined to establish a secure key between two end-users on the ground. See Figure~\ref{fig2} for a sketch of the required steps.

\paragraph*{Information-theoretic key relaying with TNs}
A quantum link at a time is established between a satellite and a target OGS, which serves one of the two end-users. The service can be provided to all locations on Earth to which the satellite can establish a direct line of sight (and above an elevation angle $\epsilon_\text{min}$) during its orbit; depending on the orbit inclination and altitude, this may encompass all of Earth's surface, or exclude the (scarcely populated) polar regions. If successful, the QKD protocol results in the generation of a secure key which is then stored in the local KMS, one located onboard the satellite and the other at the end-user. From an application layer perspective, the functionality of a QKD protocol is symmetric (two identical keys are created at both endpoints of a QKD link). Thus there is no distinction between uploading and downloading and we will use these terms interchangeably here.

Once both Alice's key $k_\upA$ and Bob's key $k_\upB$ have been generated and stored in the satellite KMS, the keys can be combined together. Upon request from Alice and Bob, the satellite broadcasts over the classical authenticated channel (which can be assumed to be public) the value $k_{AB} = k_\upA \oplus k_\upB$. This operation can be interpreted as using $k_\upB$ as a key to perform a OTP encryption of the value $k_\upA$.
Bob can then use the public value $k_{AB}$ and the secure key $k_\upB$ to compute $k_\upB \oplus k_{AB} = k_\upA$, thus securely downloading Alice's key. At the end of the procedure, the satellite erases the values $k_\upA$ and $k_\upB$ from the KMS, in order to minimise the potential surface of attack. 

\paragraph*{Architectural advantages for multi-user TN networks}
Many architectural improvements are possible over the basic TN-SatQKD just presented. Since a direct line of sight between an OGS and a LEO satellite is available only a few minutes per day and, furthermore, the optical link is blocked in the presence of clouds and (realistically) of sunlight, it may require several days or weeks to upload a key. Therefore, it is of paramount importance that the keys are generated in advance from their employment and stored in the KMS, so that they are immediately available upon user request, as we explain now. 

The key combination and broadcasting step only involves the transmission of a small amount of classical data, which can be done in all weather and illumination conditions over a Radio-Frequency (RF) channel. By exploiting already existing satellite telecommunication infrastructure, the communication round (from the initial key request and to the final acknowledgement messages) can be completed with sub-second total latency from anywhere in the world, while the availability can be in the triple-nine regime. This is particularly important in the practically relevant case where a multitude of end-users are present, but a user may not know in advance with which other users she may want to communicate privately; i.e., new pairs of end-users that issue the delivery of a secure key to them can dynamically appear. In such case, each user can continuously upload secure keys on the satellite, even months or years prior to the moment in which the key will be employed. When a new end-user pair jointly issues a QKD key, this can be created (by combining two pre-existing keys) and relayed via RF in real time.

Also note that in the key combination step half of the key material is discarded. In more complex scenarios this overhead can be computed via network coding approaches\cite{bassoli2013network}. These include, for instance, the distribution of the same secure key to three or more end-users, a.k.a.\ the conference key agreement problem; or the case in which the keys obtained from multiple independent SatQKD providers are combined so that the key is compromised only if all the providers are malicious, thus obtaining security enhancements.

\paragraph*{Key relaying without information-theoretic security}
In alternative, one could abandon the information-security framework for the sake of efficiency in the key relay process. For example, suppose Alice has uploaded a long key $k_\upA$ (e.g., 1 megabyte) and Bob a relatively short key $k_\upB$ (e.g., 1 kilobyte). On the satellite Bob's key can be used to encrypt Alice's key with a strong symmetric encryption algorithm (e.g., with the 1024-bit version of AES), broadcast the resulting ciphertext, which Bob can then decrypt using the local copy of $k_\upB$ and thus securely retrieve $k_\upA$. In this approach an arbitrarily small fraction of key material has to be sacrificed for the secure key relaying ($0.1\%$ in this example).

\paragraph*{Satellite networks consisting of both TN and UTN}
In principle, it may be possible to employ multi-hop networks mixing both TN and UTN nodes. The secuirty of such link architectures is lower than those that avoid the use of TN altogether, which is the main appeal of using the complex protocols such as EB-QKD and MDI-QKD. A possibility could be to employ UTN on the ground and TN on the satellites, as the latter should be difficult to be physically accessed by a potential eavesdropper. We leave the investigation of mixed TN/UTN networks to future work.

\begin{tcolorbox}{%
\paragraph*{Conclusions}
Using a single satellite as TN it is possible to establish a QKD key between (almost) any pair of locations on Earth. This has architectural advantages, especially in multi-user scenarios where the end-user pairs are dynamically defined and the QKD key has to be relayed with sub-second latency and high reliability.
}
\end{tcolorbox}

\subsection{Selection of the quantum link direction (DL, UL)}
\label{sec:DL}

SatQKD protocols implementation requires an FSO quantum link, which may be established in DL direction (space to ground) or in UL direction (ground to space). The option with better coupling efficiency between Alice and Bob as resulting from the quantum link budget shall be selected.
The requirements of classical FSO links only marginally influence the choice between DL and UL since some functionalities, such as classical communication and PAT, typically require bidirectional links anyway.

\paragraph*{Trade-off criteria}
The trade-off criteria here are the quantum link budget parameters of the atmospheric free-space channel that depend on the link direction. Therefore, this excludes static parameters like antenna gain, insertion losses and atmospheric extinction which are independent on link direction. It is possible that specific elements may have different specifications in its space or ground version. In this study it is assumed that individual elements and subsystems have same specifications in space and ground, e.g. insertion loss of an optical system is assumed to same in space and on ground. 
The resulting trade-off criteria are then the intensity scintillation, the wave-front perturbations, beam wandering and beam broadening. All these are effects due to atmospheric turbulence. The asymmetry stems from the fact that turbulence is present only close to Earth's surface and therefore, the signal is perturbed only in the last few kilometers, closest to the OGS. Furthermore, certain technological implications resulting from these effects are explained and assessed in the next paragraphs. 

\paragraph*{Intensity scintillation}
The propagation of an optical signal through the atmospheric turbulence results in intensity scintillation of the beam. Origin of this scintillation are wave-front distortions and resulting self-interference of the beam after certain propagation distance. The signal scintillation can easily exceed dynamics in the order of \SI{10}{dB} over millisecond timescales and appears in UL and DL direction. Aperture-averaging reduces the scintillation for cases when the coherence width of the intensity field is smaller than the receiving aperture. Since intensity scintillation is not an extinction of the signal but a redistribution of energy due to interference, it has no impact on mean end-to-end transmission efficiency of the quantum link in either direction. 
However, scintillation can still cause a decrease of the received quantum signal if the click rate depends non-linearly on the end-to-end transmittivity, for instance when the system operates close to the saturation regime determined by the dead-time of the employed single photon detectors\cite{scriminich2022optimal} or in CV systems\cite{dequal2021feasibility}. Nonetheless, these non-linear effects are expected to be negligible in most SatQKD systems because the system's operation point is set far away from the saturation regime. In summary, scintillation's impact is often negligible in both up- and downlink direction and, thus, here not considered in the UL/DL trade-off.

\paragraph*{Wave-front distortions}
For DL quantum channel the main atmospheric turbulence effects decreasing the link transmission are wavefront distortions. These can be expanded in terms of Zernike modes. The zeroth-order mode describes a global phase shift of the wavefront (a.k.a.\ phase piston), first-order modes describe Angle-of-Arrival (AoA) fluctuations and higher-order modes describe different kinds of wave-front distortions\cite{noll1976zernike}. The phase piston only affects QKD protocols in which a global phase reference has to be maintained, such as CV-QKD or TF-QKD, and is symmetric in UL and DL. First-order Zernike modes give the Angle-of-Arrival (AoA) fluctuations. AoA fluctuations are typically very well corrected by the fine-tracking subsystem, which is part of the PAT systems. Consequently, only higher-order wavefront perturbations differently affect the UL and DL channels.

For the DL channel, wave-front distortions can significantly worsen the focusing capability of the signal. In turn, focusing limits the achievable efficiency in coupling the signal to the detector. The length scale of wave-front distortions are typically expressed by the so-called Fried parameter\cite{fried1966optical}. Compared to the telescope aperture, a small Fried parameter denotes strong distortions, large Fried parameters weak distortions. These distortions can be mitigated by deployment of an AO system which corrects the wave-front and thus, increases focusing capability for the signal again. Alternatively, the use of free-space detectors with larger detector areas (compared to the mode field of the fiber) might render an adaptive optics system unnecessary.  

For the UL channel wavefront distortions have a negligible impact on focusing of the signal within the satellite receiver. The wavefront accumulates distortions only while propagating through the atmosphere. After that the perturbations are almost frozen in shape and are stretched-out as the beam expands (due to beam divergence) while propagating towards the satellite. More precisely, Fried parameter in UL scales as $O(L^{11/10})$, where $L$ is the link distance\cite{andrews2005laser}. For LEO satellites, the UL Fried parameter is of the order of several meters, much larger than the size of any satellite-mounted telescope. As a consequence, almost diffraction-limited focusing within the satellite is possible, even without the use of AO systems.

\paragraph*{Beam wandering and beam broadening}
For the UL quantum channel the main atmospheric turbulence effect increasing the mean link loss is the increased long term spot size of the transmitted laser beam due to beam broadening. The long term spot size is caused in combination by diffraction of the unperturbed beam, a short-term beam spread due to scintillation and the beam wander due to large scale turbulence cells\cite{andrews2005laser, dios2004scintillation}. The diffraction of the unperturbed beam can easily be controlled by selecting the transmit aperture and the short-term beam spread can often be neglected compared to the impact of the beam wander on the long-term spot size. The beam wander cannot be sufficiently mitigated in a UL configuration by employing pointing-by-tracking beam stabilisation, as typically done. Reason is the PAA being much larger than the isoplanatic angle and therefore, the DL signal does not serve as a good reference signal for beam stabilization. Therefore, the LEO uplink beam wander cannot be compensated which results in a still strong long-term spot size (beam broadening) and thus, signal loss.  

For the DL channel the beam spreading and beam wandering are negligible. In fact, when the DL beam reaches the top of the atmosphere the transversal size of the beam is already tens of meters in size and only a few kilometers away from the OGS. Atmospheric turbulence can then increase beam divergence and pointing jitter by tens of microradians, which then shifts and broadens the beam spot at the OGS by a few more centimeters.

\paragraph*{Technological trade-offs}
In DL a main aspect that hinders link efficiency is coupling the received optical signal to the quantum detector. This is made more challenging by wavefront perturbations due to the presence of atmospheric turbulence. The trade-off is different for free-space-coupled detectors and fibre-coupled detectors. Free-space coupled detectors such as single photon avalanche diodes have detectors sizes in the range of few ten \SI{50}{\micro\meter}. In contrast, a system coupled to a Single-Mode Fibre (SMF) has to focus the light onto the SMF tip. The field mode diameter of a standard SMF-28 is around \SI{10}{\micro\meter} in the C-band. AO systems may be needed to couple the signal into the fiber with the desired efficiency or, in alternative, large  signal coupling losses has to be taken into account. Another technological possibility is to couple into a Multi-Mode Fibre (MMF), which have a diameter larger than that of an SMF and allows mitigation of the coupling losses. The mode dispersion in a MMF allows coherently transporting the signals only for few tens of meters, but that is sufficient for placing the detector in a separated location from the telescope.

For UL the main challenge is the correction of beam wandering. A strategy for partial correction of beam wandering is that of pointing the UL beam in the direction from which the DL beacon is received. However, this is not sufficient to fully correct the beam wander, since the UL and DL atmospheric channels are spatially separated by a quantity given by the satellite PAA. For LEO links the PAA typically exceeds the isoplanatic angle, meaning that the light in UL effectively passes through different turbulence eddies and thus is subject to different beam refraction\cite{andrews2005laser}. Better correction performances could be obtained employing an artificial star guide system to have a better estimation of the atmospheric turbulence. The guide star can be placed in the need position ahead the satellite tracking to receive the correct beacon signal for pointing-by-tracking. The Technology Readiness Level (TRL) of this technology is however rather low and its performance is still not as good as that of AO systems for DL signals\cite{esposito_focus_1996, mata-calvo_laser_2017}.

\begin{tcolorbox}{%
\paragraph*{Conclusions}
Beam wandering and wave-front distortions are dual phenomena for the UL and DL channels, the former being relevant only in UL and the latter only in DL. The losses due to beam wandering (and thus strong long-term beam broadening) in UL cannot be compensated to a satisfactory degree because of the point ahead angle. The losses due to wave-front distortions in the DL can either be compensated to a large degree with adaptive optics systems or can be reduced/minimized by using free-space detectors with larger chip sizes. These considerations result in DL being currently more favourable than UL for SatQKD implementations.
}
\end{tcolorbox}

\subsection{Selection of discrete-variable or continuous-variable protocols (DV, CV)}
\label{sec:DV}

Two broad families of QKD protocols can be identified, depending on how the quantum information is encoded: Continuous-Variable (CV) and Discrete-Variable (DV) protocols. The first involves quantum information encoded in an infinite dimensional Hilbert space. The second involves quantum information encoded in a finite-dimensional Hilbert space, typically a qubit, i.e., a two-dimensional quantum system\cite{pirandola2020advances}. 

The boundary between CV and DV protocols is nuanced, since DV protocols can be implemented with WCP, i.e., coherent states that are defined in an infinite-dimensional Hilbert space. Thus, a protocol is defined as DV if one can identify a finite-dimensional Hilbert subspace in which the quantum information is encoded, while the orthogonal subspace only encode side-channel (spurious) information. The QKD security proof has thus to include a reduction from the real, physical implementation of the QKD protocol to the ideal, mathematical description of the finite-dimensional QKD protocol. 

\paragraph*{Continuous-variable protocols}
In CV quantum information encoding the employed Hilbert space is the one associated to one or more harmonic oscillators. These are in practice always realised as modes of an electromagnetic field, e.g., as a coherent laser signal. For a single electromagnetic field mode one can identify two quadratures, customarily denoted as $x$ and $p$ in the physics literature and which are the (non-commutative) quantization of the $I$ and $Q$ quadratures known in classical coherent communication. The meaning of these quadratures is most straightforwardly defined  operationally in terms of the physical setting required to measure them. This can be accomplished by means of homodyne detection. In an homodyne detector the electromagnetic signal is impinged on a $50:50$ beam splitter, while on the other input port is coupled a strong reference electromagnetic signal, the so-called Local Oscillator; the signals at the output ports are then measured by two linear-response photodetectors and the difference of the two photocurrents provides, by definition, a measurement of the $x$ quadrature of the input signal. The $p$ quadrature is obtained if the phase of the LO is shifted by $\frac{\pi}{2}$ compared to the $x$ quadrature measurement.

Thanks to theoretical advancement over the course of 25 years, CV-QKD protocols have reached a high level of maturity. The first CV-QKD protocols required entanglement\cite{ralph1999continuous} or squeezing\cite{hillery2000quantum}; it was later realised that coherent-states (that is, WCP) are sufficient to implement a CV-QKD protocols\cite{grosshans2002continuous}. This protocol could only tolerate at most \SI{3}{dB} of loss, but this restriction was later overcome by means of reverse information reconciliation, i.e., by using Bob's signal as reference, rather than Alice's\cite{grosshans2002reverse}. Security proofs were also strengthened, showing first that security holds against collective attacks\cite{navascues2006optimality} and then proven against fully general attacks\cite{renner2009finetti}. More recent development include the development of secure discretely modulated CV-QKD protocols (which is more practical than continuous Gaussian modulation)\cite{lin2019asymptotic} and the use of the entropy accumulation theorems\cite{dupuis2020entropy, metger2022generalised} to yield tight end-to-end SKL bounds against general attacks\cite{jain2022practical}.

CV-QKD features some advantages compared to DV-QKD, stemming from the employment of homodyne receivers instead of single-photon detectors. A very narrow-band spectral filtering is naturally applied, essentially given the requirement of matching the LO spectrum. This stems from the fact that the measured photocurrent signal is extracted by the interference between the input optical signal and the LO; any signal out of band signal will result in fast oscillations that are then averaged out. This can be beneficial in FSO links, since they typically feature higher rates of BackGround-Light (BGL) than fibre-based links. Furthermore, homodyne receivers very high signal detection rates at a fraction of the cost of typical single photon detectors in the C-band. The potential cost savings are however not so compelling in SatQKD, which in any case requires a great deal of custom equipment.

Albeit several feasibility studies of CV protocols in SatQKD have been put forward\cite{he2019photonic, pan2020security, zuo2020atmospheric, xu2021noiseless, dequal2021feasibility}, these do present important shortcomings that make them more challenging to employ than DV ones. A first shortcoming is the higher susceptibility to power scintillation\cite{dequal2021feasibility}. Secondly, the LO frequency (if generated within Bob's receiver) has to compensate the Doppler shift due to the satellite change in velocity. Finally, and most importantly, CV-QKD are significantly less robust to transmission losses, since the optical signal detection of is based on the measurement of a small deviation above the shot-noise limited interferometric measurement with the LO. In practice, it means that with currently existing homodyne detectors positive SKR can only be achieved for losses not exceeding circa $20 - \SI{25}{dB}$\cite{wang2019realistic}, which is very challenging to achieve in satellite-to-ground links. In comparison, positive SKR with channel losses in excess of $\SI{69}{dB}$ (in fibre) have been experimentally demonstrated with the DV-QKD, namely, with the BB84 protocol\cite{boaron2018secure}.

\paragraph*{Discrete-variable protocols}
DV-QKD protocols have a comparatively high level of maturity. Methods for dealing with imperfections in the physical device implementations have been known for more than 20 years\cite{gottesman2004security} and the first security proofs against fully general attacks have arrived shortly after\cite{renner2008security}. 

Among these PM-DV-QKD protocols, BB84 protocol stands out as being the most promising candidate. Its combines a rather simple implementation (compared, e.g., to high dimensional protocols) with well vetted security, both in terms of security proofs and of practical implementation. On the theory side, it features linear scaling of the key rate in the channel transmission, $R = O(\avg{\eta})$; this is true both when using SPS and WCP, the latter thanks to the employment of the decoy-state method\cite{lo2005decoy, lim2014concise}.

\begin{tcolorbox}{%
\paragraph*{Conclusions}
DV-QKD protocols are favoured compared to CV-QKD protocols mainly due to their ability of achieving a positive key rate at significantly higher transmission losses ($20-\SI{25}{dB}$ in one case, $60-\SI{70}{dB}$ in the other). This, in turn, significantly eases the requirements on the optical link budget. 
}
\end{tcolorbox}

\subsection{Selection of photon source technology (SPS, WCP)}
\label{subsec:SPS_WCP}

SPS would be the first choice as DV-QKD, in the idealised formulation, is based on the assumption that single-photon states (i.e., single qubits) are being sent. Conversely, the use of WCP  entails the preparation of states with multi-photon components, which is not immediately compatible with the security proofs of ideal QKD. However, since 2005\cite{lo2005decoy} it is known that employing pulses with lower intensity, the so-called decoy states, enables the possibility to extract a secure key at relatively high rate, i.e.\ scaling linearly in the end-to-end transmission efficiency $\eta$. The security proof of the decoy-state method has been thoroughly examined and vetted\cite{lim2014concise, trushechkin2021security}. The fundamental step is to provide a lower bound on the number of single-photon events that resulted in a click at the QKD receiver; through the use of privacy amplification, these events allow extracting a final secure key\cite{trushechkin2021security}.

\paragraph*{Single-photon sources}
Several platforms are currently used to generate single photons\cite{Cao.2019}, such as spontaneous parametric down-conversion\cite{Bock.2016, Yin.2020}, spontaneous four-wave mixing\cite{wakabayashi2015time}, carbon nanotubes\cite{He.2017} or atomic sources\cite{Willis.2011}. 

The most promising technology for good-quality, high-performance SPS is the fabrication of semiconductor Quantum Dots (QD)\cite{Vajner.2022}. First developed in 1993\cite{Leonard.1993}, quantum dots are sometimes referred to as artificial atoms, as their energy bands are discrete and behave like atomic energy levels. When excited, the QD promotes an electron to its conduction band which fills up like an atomic s-shell. The decay of the electron will hence generate only one photon at a time\cite{Michler.2000}. The first QKD experiment using QD was carried out in 2002\cite{Waks.2002}. Since then, the field has seen many improvements, with QD-based sources being used to improve the performance of QKD links both for PM protocols\cite{Intallura.2009, Heindel.2012, bozzio2022enhancing} and in EB implementations\cite{Dzurnak.2015, Schimpf.2021}.

Notwithstanding the several advances that have been made in the last years, which have drastically improved the performance of SPS-based QKD, current implementations of single photon sources are not competitive with laser-based sources. The main drawback of this solution is that repetition rates are orders of magnitude lower than what a standard QKD transmitter can produce\cite{Takemoto.2010, yang2023high},  which is aggravated by the fact that these sources are non-deterministic, typically emitting a photon with 10\% probability or less. Finally, the fabrication TRL of these devices is rather low, especially in the C-band, and is not yet at the point where they can be taken out of the lab for field trials and commercial applications. For instance, these devices typically operate at cryogenic temperatures and the use of cryostats is hardly compatible with the SWaP of satellite platforms.

\paragraph*{Weak coherent pulses}
The generation of WCP is based on the use of laser technology and of (variable) optical attenuators, components having very high repetition rates an very high TRL. A further advantage of using WCP is that optical transmission losses within the transmitter terminal (e.g., due to finite reflectivity of the mirrors and of the lenses) are irrelevant: in any case the initial laser signal has to be strongly attenuated (typically by $60 - \SI{90}{dB}$) and the optical transmission losses can be measured and compensated in such a way that the photon intensity at the external aperture of the LCT reaches the target value. 

One complication of the use of WCP is that each WCP sent requires a phase randomisation, so that the employed photonic state is indistinguishable from a statistical mixture of states with a defined photon number. This can be achieved by active randomisation, via a phase modulator\cite{zhao2007experimental}, or exploiting the phase randomness of the spontaneous photon emission initiating a lasing event\cite{yuan2016directly}. The second disadvantage is that a decoy-state QKD protocol yields a lower key rate compared to one employing a similarly-performing (in terms of fidelity and generation rate) single-photon source, typically by about one order of magnitude. 

It is also possible to realise WCP via a completely passive scheme, avoiding the use of active modulation elements\cite{wang2023fully, zapatero2023fully}. This has security advantages as these schemes are inherently immune from side-channel attacks on the modulator component, such as Trojan horse attacks\cite{vakhitov2001large}. The downside is that  the quantum state preparation cannot be exact and could result in increased QBER. This approach is promising but not yet fully explored and we thus leave out of our trade-off analysis.

\begin{tcolorbox}{%
\paragraph*{Conclusions}
The use of WCP stands as a much more advantageous approach than the use of single-photon sources: laser technology has a much higher TRL than SPS, is more compatible with satellite platforms, the secure key generation rates are much higher under realistic system performance assumptions. The security proofs when WCP are used in conjunction with the decoy state method are very mature.
}
\end{tcolorbox}

\subsection{Selection of a reference protocol within the decoy-state PM-DV-QKD class (BB84)}

We now arrive at the selection of the specific protocols that offer the best overall performance within the class identified in the previous sections. The main desiderata the protocols should have are: (1) a high key-generation rate, with linear scaling in the quantum channel transmission, $R = O(\eta)$; (2) a high maturity of the implementation security, including at least existence a end-to-end proof of security against coherent attacks and of a squashing model for the receiver.

\paragraph*{High dimensional protocols}
For High-Dimensional (HD) protocols we here loosely group QKD protocols where information is encoded in finite Hilbert spaces but its dimension is a parameter of the QKD protocol that can, in principle, be made arbitrarily large. In some HD protocols a single photon detection can have $d$-ary outcomes, rather than a binary outcomes, so that each photon can asymptotically contribute $\log_2(d)$ bits of randomness in the generation of the QKD key; this results in higher information capacity. In other protocols, the measurement outcomes remain binary, but the number of possible measurement bases grows with the Hilbert space dimension; this can provide better resilience against noise.

Some early proposals of HD-QKD protocols that have sparked significant interest were the Differential Phase-Shift (DPS) protocol\cite{inoue2002differential} and the Coherent One-Way (COW) protocol\cite{stucki2005fast}. These were conceived because of the implementation simplicity: in their simplest realisation they only require a pulsed laser source, together with a single phase modulator (for DPS) or a single amplitude modulator (for COW). For several years it has been conjectured that full security proofs against coherent attacks could be given for these protocols. However, a recent result has shown the existence of zero-error attacks against COW that very severely reduce the achievable key rate\cite{trenyi2021zero}. Recently, security against coherent attacks have been given for variants of DPS\cite{mizutani2023finite} and COW\cite{gao2022simple}, but with a SKR scaling only quadratically in the quantum channel transmission, $R=O(\eta^2)$.

Another HD-QKD showcasing very interesting properties is the so-called Round-Robin (RR) protocol\cite{sasaki2014practical}. The QKD transmitter encodes a sequence of bits onto non-orthogonal quantum states and the QKD receiver randomly selects which bit should be extracted; by the complementarity principle, this bit cannot be reliably obtained by Eve. Interestingly, the RR protocol with WCP achieves a linear scaling of the SKR in the channel transmission, without the need of using decoy states. The maturity of the security proof is also increasing as showcased, e.g., by new tight finite-key analyses\cite{liu2021tight}.

Many other HD-QKD protocols exist\cite{otte2020high} and it seems that some could be competitive compared to qubit-based protocols, especially in the high-QBER regime. However, the comparison of noise-resilience is non-tirvial and model-dependent, as the increased number modes often results in a larger coupling to BGL and, thus, increase in QBER. And even if high dimensional protocols were superior in terms of noise resilience or channel capacity, it is unclear if this is sufficient to justify the corresponding increase in transmitter and receiver complexity\cite{cozzolino2019high}.

\paragraph*{BB84 and closely related protocols}
Qubit-based protocols which are the simplest and thus are, in many application scenarios, the favoured choice. These can be roughly classified by the number of states that can be prepared by the transmitter and measured by the receiver. We consider here only a few qubit-based QKD protocols, which have been arbitrarily selected. Namely, these are the BB84 protocol (which employs the eigenstates of the Pauli-X and Pauli-Z basis)\cite{bennett1984quantum}, the 6-state protocol by Bru{\ss} (which employs the eigenstates of the Pauli-X, Pauli-Y, and Pauli-Z basis)\cite{bruss1998optimal} and the 3-state protocol by Rusca (which employs the eigenstates of the Pauli-Z basis and only the positive eigenvalue of the Pauli-X basis)\cite{rusca2018security}. Many other protocols, such es e.g.\ the one introduced by Scarani, Acín, Ribordy, and Gisin is a 2004 (SARG04)\cite{scarani2004quantum}, are discarded because have a sub-linear SKR scaling in the channel transmission when implemented with WCP.

The 6-state protocol is a rather natural extension of the BB84 protocol. One main advantage is that, due to three measurement bases and thus better quantum state estimation, allows more noise tolerance assuming a depolarising channel\cite{pirandola2020advances}. Surprisingly, however, no security proof exists yet that provides finite-key analysis for the decoy-state version of the protocol. Similar results have been available for BB84 for more than ten years\cite{lim2014concise}. 

The main idea of the 3-state protocol is to exploit the fact that only the Pauli-Z basis is employed to generate a key, while the Pauli-X basis is only used to monitor the QBER. This, combined with other ideas that allow to increase the pulse repetition rate rate\cite{rusca2018security}, typically results in a higher SKR than BB84 for caomparable transmitter and receiver technologies. However, the security proof is rather incomplete, lacking a proof against general attacks and, furthermore, no detector squashing models has been found yet.

BB84, in contrast, has been very thoroughly investigated and optimised over the years by many experimental and theoretical research groups. As an example, a well-known optimisation method is to increase the probability of choosing the Pauli-Z basis and decreasing the Pauli-X basis for both parties; in such case only the Z-basis is used for key generation and the X-basis for signal monitoring. This results in a higher sifting rate and, downstream, a higher SKR\cite{lo2005efficient}. Furthermore, BB84 inching towards having security analyses capable of addressing all the implementation loopholes a the same time. This has been demonstrated, for instance, by a recent in-depth investigation of the commercial BB84 implementation by the company QRate\cite{makarov2023preparing}.

\paragraph*{Reference-frame independent protocols}
A tantalising possibility that could seem particularly suited to the SatQKD scenario would be the use of a Reference-Frame-Independent (RFI) QKD protocol\cite{laing2010reference}. This allows performing QKD even in absence of a common reference frame (i.e., a system of coordinates) between Alice and Bob, which allows them to consistently interpret the information encoded in a transmitted qubit. In practice, this would allow foregoing implementing dedicated subroutines for reference frame alignment. While the original protocol relied on entangled states, a simpler RFI protocol only requiring separable qubit states can be devised if Alice's and Bob's apparatuses share at least one common axis of reference\cite{wabnig2013demonstration}. In polarisation-encoded SatQKD that is the case: the common axis is given by the line-of-sight connecting their terminals.\footnote{A further assumption, which is in practice satisfied by atmospheric channels, is that birefringence should be negligible.}

Unfortunately, existing RFI-QKD protocols are ultimately not applicable to SatQKD since the relative orientation between Alice's and Bob's reference frames, albeit unknown, has to remain fix. This is not the case in dynamic satellite links, where the attitude of the satellite and ground terminals change over time. We thus restrict our analysis to DV-QKD protocols whereby Alice's and Bob's reference frames are assumed to be (approximately) aligned.

\begin{tcolorbox}{%
\paragraph*{Conclusions}
The decoy-state version of BB84 has the highest maturity in terms of implementation security among PM-DV-QKD protocols, rendering it the currently preferred choice to be employed in real-world applications. Some PM-DV-QKD protocols may have similar or even better performances in some application scenarios, but their implementation security currently lags behind  significantly. 
}
\end{tcolorbox}

\section{Comparison of some selected SatQKD implementations}
\label{sec:comparison}

Even within the quite specific SatQKD architecture defined in Section~\ref{sec:architecture} many non-trivial design choices remain to be made. Here we do not provide conclusions about which ones provide the best trade-offs, as that will depend on the specific application scenarios. The influence of the following design choices on SatQKD operations is then analysed. For sake of simplicity we here only consider the QKD established between a satellite and an OGS, while real SatQKD systems will entail links among multiple users and multiple satellites.
\begin{tcolorbox}{%
\begin{enumerate}
    \item The quantum signal wavelength, where the considered alternatives are the C-band and the Silicon band.
    \item The detector technology, where the main contenders are Single-Photon Avalanche Photo-Diodes (SPAD) and Superconducting Nanowire Single-Photon Detectors (SNSPD).
    \item The orbit parameters, including the satellite altitude and orbital plane inclination.
    \item The transmitter LCT and receiver OGS parameters, entering in the calculation of the optical link budget.     
    \item The physical modulation of the quantum signal, which is either polarisation encoding or time-bin encoding.
    \item The optimisation of the BB84 protocol parameters.
\end{enumerate}
}
\end{tcolorbox}

\subsection{Wavelength for quantum communication}

An important non-trivial trade-off consists in the choice of wavelengths that implement the quantum channel. There are two reasonable options that can be employed to this end: the C-band (around $\SI{1550}{nm}$) and the Si-band (around $\SI{850}{nm}$). 

\paragraph*{C-band}
The C-band is the wavelength band having the lowest loss in fibre (around \SI{0.18}{dB\per\kilo\meter}), allowing the propagation of optical signals over relatively long distances. Consequently, the C-band is also the wavelength for which most classical communication technology has been developed. This allows for lower cost implementation of Alice and Bob modules due to wide availability of core components, such as sources, detectors, and network components like multiplexers and demultiplexers. Furthermore, combination of the quantum and data channels in an FSO link is straightforward, since \SI{1550}{\nano\meter} is often already used for classical communication, avoiding the need to implement multi-chromatic systems.

The C-band is also advantageous as the atmosphere features a good transparency window in that spectral region. For instance, the atmospheric transmission loss from sea level to zenith is only $\SI{0.4}{dB}$ for a nominal horizontal visibility of $\SI{23}{\kilo\meter}$.

A strong architectural advantage of the C-band is that it allows a straightforward interfacing to a \textit{fiber network}, which allows routing the quantum signal to end-users that are not co-located at the OGS. This can be employed, for instance, in a so-called \textit{provider-OGS} integration concept\cite{OGS}. In such integration concept the quantum signal received from the satellite is not detected at the OGS, but it is forwarded to the end-user, who detects it with a QKD receiver module.\footnote{Note that the OGS, in this concept, does not need to be trusted by the end-user. It has no information about the quantum signals that are passing through it and effectively only a part of the quantum channel.} The low fibre losses in the C-band allow delivering the quantum signals to end-users located within a few kilometers with only a few of dB of losses. If positive SKR can still be achieved this would allow, e.g., a single provider-OGS to service all the end-users located within a city.

\paragraph*{Si-band}
The Si-band has some advantages and disadvantages compared to the C-band. The optical transmitter system complexity will typically be higher: assuming that the C-band is employed for PAT and classical data links, poly-chromatic optical system will be needed. The atmosphere is slightly less transparent in this band, having around $\SI{0.9}{dB}$ of losses to zenith at $\SI{23}{\kilo\meter}$ of horizontal visibility.

A stark potential advantage derives from the possibility of increasing the transmitter antenna gain, compared to an equally-sized terminal operating in the C-band. The diffraction-limited beam divergence at wavelength $\lambda$ for a LCT of size $D_\Tx$ is $\theta = O(\lambda/D_\Tx)$. This results in a theoretical increase of transmitter antenna gain of about $10 \log_{10}\!\left[\left(\frac{\SI{1550}{\nano\meter}}{\SI{850}{\nano\meter}}\right)^2\right] = \SI{5.2}{dB}$ for a terminal of fixed size. This allows for a better utilization of the tight SWaP of small satellite missions. Naturally, machining and polishing of the mirror surfaces to sufficient precision is more challenging for systems working at shorter wavelengths.

A disadvantage of the Si-band is that it does not allow a straightforward network integration. At this wavelength the photons will incur losses of around $\SI{3}{dB \per \kilo\meter}$ along the fibre, severely limiting the connection range. An alternative could be to use a complex wavelength transduction system to convert the FSO photons to the C-band. However, other OGS integration concepts, whereby the signal is directly detected at the OGS, are not affected by this limitation. For instance, this happens when the end-user QKD receiver and KMS are co-located with the OGS, or when the OGS is employed as a TN.

Another challenge is that shorter wavelengths are more susceptible to atmospheric turbulence since, fixing the atmospheric turbulence conditions, the Fried parameter scales as $O(\lambda^{6/5})$. This would require complex, high-order AO to approach a diffraction-limited focusing of the signal. However, Si-band signals lend themselves to being detected with large-area free-space Silicon detectors; in this case the focusing requirements to collect most of the signal are much less stringent. The choice of the quantum channel wavelength is thus intertwined with the technologies for single-photon detection available at the different wavelengths, as discussed below.

\subsection{Single-photon detector technologies}

The three technologies for single-photon detection we consider here are SNSPD, InGaAs SPAD and Silicon SPAD. A small survey of high-performance commercially available detectors (with one representative for each of these technologies) is given in Table~\ref{tab:Detector_Specs}. Other technologies do exist, including legacy photo-multiplier tubes, frequency up-conversion techniques, systems based on quantum dots, or transition-edge sensors\cite{hadfield2009single}. However, these alternatives either do not meet the performance requirements for SatQKD applications, or have too low TRL and are not readily commercially available and will not be considered here. 

\begin{table*}[t]%
    \centering
    \begin{tabular*}{\columnwidth}{@{\extracolsep\fill}lccrrrrr}
        \toprule
        \multicolumn{1}{c}{\textbf{Detector}}       & \multicolumn{1}{c}{\textbf{Vendor}}          & 
        \multicolumn{1}{c}{\begin{tabular}{c}\textbf{Free-space} \\ \textbf{vs. SMF}\end{tabular}} & 
        \multicolumn{1}{c}{\textbf{Wavelength}}      & \multicolumn{1}{c}{\textbf{Eff.}}     & \multicolumn{1}{c}{\textbf{DCR}} & \multicolumn{1}{c}{\begin{tabular}{c}\textbf{Dead} \\ \textbf{time}\end{tabular}}      & \multicolumn{1}{c}{\textbf{Jitter}}\\
        \midrule
        SNSPD ID281 & IDQ & SMF & 
        \SI{1550}{\nano\meter}  & 0.90 & \SI{90}{\hertz} & \SI{30}{\nano\second} & \SI{30}{\pico\second} \\
        ID Qube NIR & IDQ & Free-space & 
        \SI{1550}{\nano\meter} & 0.20 & \SI{3000}{\hertz} & \SI{100}{\nano\second} & \SI{200}{\pico\second} \\
        SPCM-850-14 & Excelitas & Free-space & 
        \SI{850}{\nano\meter} & 0.58 & \SI{100}{\hertz} & \SI{22}{\nano\second} & \SI{350}{\pico\second}\\
        \bottomrule
    \end{tabular*}
    \caption{Examples of commercially available detector systems, each representative of a distinct single-photon detection technology. The first is an SNSPD; the second is a SPAD detector based on the InGaAs semiconductor; the third is a SPAD detector based on Silicon. The parameters reported here (Eff.: efficiency, DCR: dark count rate) have been employed in the SatQKD simulations below and are taken from the respective data sheets \cite{IDQ2021,IDQ2023,Technologies2024}.}
    \label{tab:Detector_Specs}
\end{table*}

\paragraph*{Superconducting Nanowire Single-Photon Detector}
SNSPD has emerged as the best technology for DV-QKD applications in the C-band. It operates just below the threshold of superconductivity, such that the energy deposited by any photon absorption breaks the superconductivity and triggers the detection signal. It features high detection efficiency ($\sim 90\%$), low dark count rates ($\sim\SI{100}{\hertz}$), low timing jitter ($\sim\SI{50}{\pico\second}$), fast recovery times ($\sim\SI{30}{\nano\second}$), and even multi-photon resolving capabilities. The longest distance fibre-based DV-QKD experiment\cite{boaron2018secure}, as well as the highest SKR demonstration\cite{li2023high}, both used a SNSPD-based QKD receiver module. The downside are high procurement costs and high system complexity, as they operate at cryogenic temperatures ($<\SI{2}{\kelvin}$). Furthermore, currently commercially available SNSPD only arrive as SMF-coupled systems. Employing an SNSPD a a free-space coupled detector would require a bespoke system development and is not considered here.

\paragraph*{InGaAS SPAD}
SPAD based on the InGaAS semiconductor is an alternative technology for single-photon detection in the C-band.  Employing a InGaAS SPAD both with free-space-coupling and fibre-coupling is possible. Compared to SNSPD, they are significantly cheaper and simpler: they only require moderate cooling ($\sim\SI{-40}{\celsius}$) for Geiger-mode operation, which can be readily achieved with a thermoelectric cooler. The downside is that all system performance parameters are significantly worse, including efficiency ($\sim \SI{20}{\percent}$), dark count rates ($\sim \SI{1}{\kilo\hertz}$), timing jitter ($\sim \SI{100}{\pico\second}$). After each detection there is a certain probability of an afterpulse occuring, i.e. charge carriers getting trapped and subsequently triggering another avalanche, leading to a correlated noise detection. This can be counter-acted by introducing a prolonged dead time, either by means of passive or active quenching, such that with high probability the trapped charge carriers are depleted. Setting a relatively short dead-time ($\SI{100}{\nano\second}$) then results in a in increase of the of the effective dark-count rate ($\sim \SI{3}{\kilo\hertz}$).

\paragraph*{Silicon SPAD}
The higher energy of photons with a wavelength around \SI{850}{\nano\meter}, compared to those in the C-band, enables easier detection by means of Silicon-based SPAD. These are a very mature and relatively inexpensive technology, showcasing good performance parameters at a fraction of the complexity required for single-photon detection in the C-band. Operations at room-temperature is possible, albeit cooling is typically needed to achieve optimal detection efficiency. Si-SPAD have rather high efficiency ($\sim 60\%$), low dark count rates ($\sim\SI{100}{\hertz}$), and rather low timing jitter ($\sim\SI{200}{\pico\second}$). Being a SPAD-based technology a detection can result in afterpulses, but employing rather short dead times ($\sim\SI{20}{\nano\second}$) is sufficient to contrast this effect.

\subsection{Satellite orbit}

\begin{figure*}[t]
\centerline{
    \includegraphics[height=150pt]{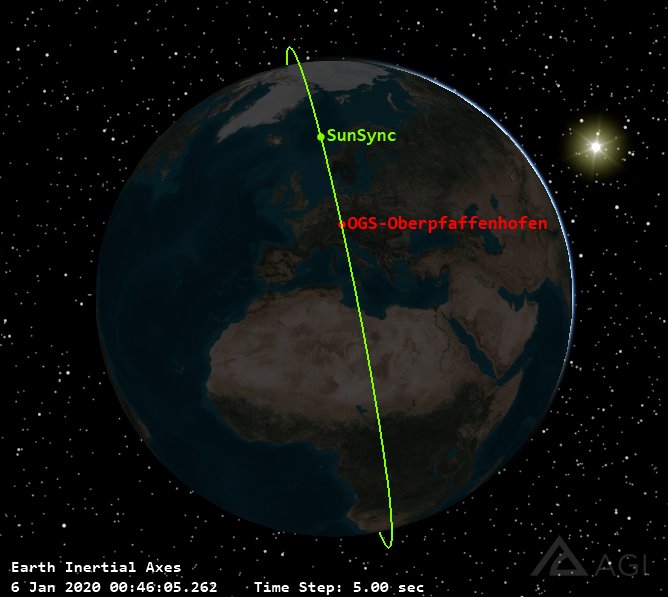}
    \includegraphics[height=150pt]{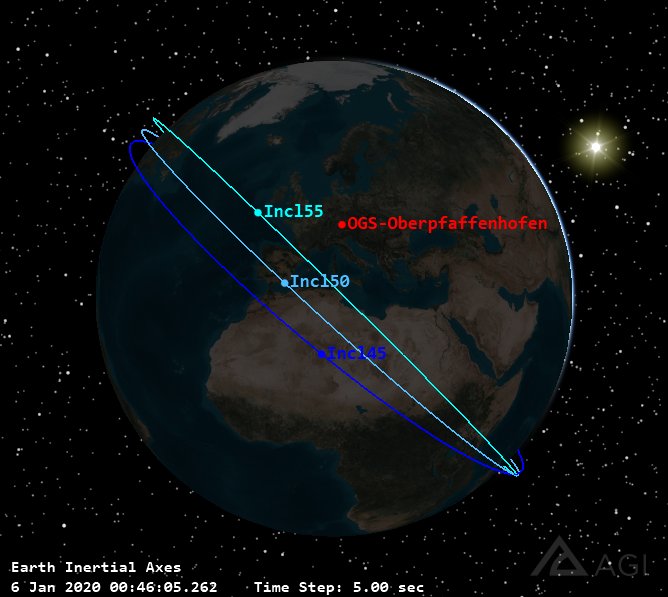}
    \hspace{-3mm}
    \raisebox{20mm}{
    \begin{tabular}{c}
    \includegraphics[height=135pt]{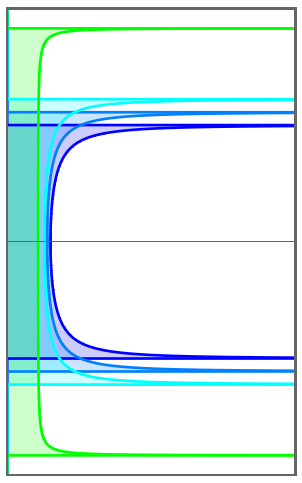}\\[10mm]
    \end{tabular}
    \hspace{-3mm}
    \begin{tabular}{l}
    \raisebox{1pt}{\textcolor{blue} {\rule{5mm}{2pt}}} $i=45^\circ$\\
    \raisebox{1pt}{\textcolor{blan} {\rule{5mm}{2pt}}} $i=50^\circ$\\
    \raisebox{1pt}{\textcolor{cyan} {\rule{5mm}{2pt}}} $i=55^\circ$\\
    \raisebox{1pt}{\textcolor{green}{\rule{5mm}{2pt}}} $i=97.66^\circ$\\[10mm]
    \end{tabular}}
    }
\caption{Illustration of some possible satellite orbits, made using the System Toolkit simulation software. On the left, a SSO is represented, while in the center three different custom-inclination orbits are shown. On the right, a plot the probability density function of the satellite latitude (obtained as a projection on Earth's surface of the satellite position in space) is given. }
\label{fig:orbits}
\end{figure*}

We here consider two classes of low-Earth orbits, categorised according to the inclination $i \in [0^\circ, 180^\circ)$: Sun-Synchronous Orbits (SSO) and custom inclination orbits. These are illustrated in Figure~\ref{fig:orbits}. 

\paragraph*{Sun-synchronous orbits}
SSO are highly-inclined slightly retrograde orbits. The quadrupolar moment of Earth's gravitational field causes a precession of the orbital plane for all inclined orbits, but in SSO the altitude-inclination relation is chosen so that the orbit precesses at a rate matching Earth's revolution around the Sun, so that the orientation with respect to the Sun is kept fixed. For circular orbits (having eccentricity $e=0$) one can find an approximate inclination of a SSO with radius $R_\textup{sat}  =  R_\oplus + h$ using the equation 
$\cos(i) \simeq -\left(\frac{R_\oplus+h} {\SI{12352}{\kilo\meter}}\right)^{7/2}$, where $h$ is the satellite altitude and $R_\oplus = \SI{6378.137}{\kilo\meter}$ is Earth's equatorial radius\cite{wertz1999space}. By choosing an altitude of $h = \SI{567}{\kilo\meter}$ the orbital period is \SI{96}{\minute}, which leads to exactly 15 satellite orbits per solar day, with a corresponding inclination of $i= 97.66^\circ$. Next, a RAAN is selected so that one satellite pass for a given OGS location occurs in the middle of the night. This guarantees one satellite pass each night, happening always at the same hour, all year round, and all having similar elevation-over-time link dynamics. This may be convenient for SatQKD experimental demonstrators, as it allows consistent conditions for experimentation. 

\begin{figure}
    \centering
    \includegraphics{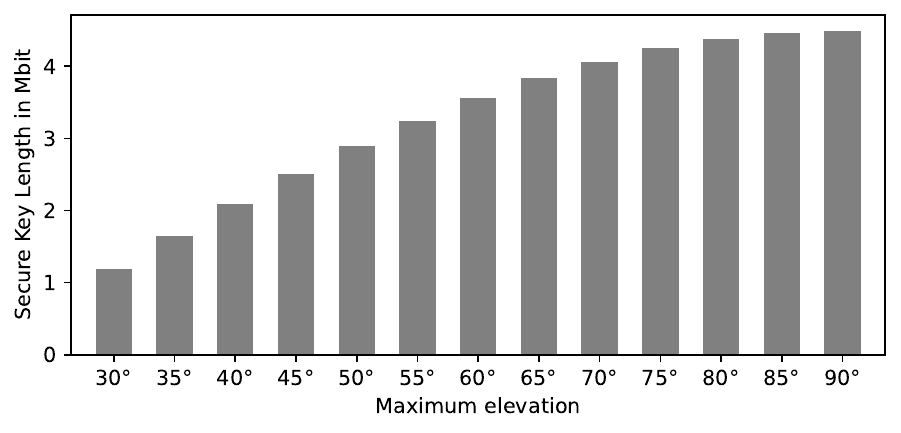}
    \caption{Secure key length obtained by satellite passes with different maximum elevations and a fixed minimum elevation of $\SI{20}{\degree}$ for the reference SatQKD system.}
    \label{fig:SKL_maxElev}
\end{figure}

We have simulated a reference SSO satellite pass, that we have then employed for the analyses in the rest of this Section. To so, we have provided the initial conditions to the orbit propagator VENQS, developed at the Institute for Satellite Geodesy and Inertial Sensing of the German Aerospace Center (Deutsches Zentrum für Luft- und Raumfahrt, DLR). The satellite state vector, consisting of the position and attitude, is simulated for each time step with a one second resolution. Knowing the satellite and OGS location allows to calculate the line of sight vector, from which the link range and elevation angle above the OGS local horizon are computed. This is illustrated in the top-left part of Figure~\ref{fig:LinkBudget}. The maximum elevation angle of this orbit is here chosen to be $80^\circ$, since this is the maximum elevation that can be reliably tracked employing a two-axes mounted telescopes: the pointing control in azimuth and elevation results in a singularity of the control coordinates at the zenith, requiring very fast azimuthal rotation to track satellite passing near that point. However, as displayed in Figure~\ref{fig:SKL_maxElev}, have also simulated the performance of a reference QKD protocol for satellite passes featuring different maximum elevations, ranging from $\SIrange{30}{90}{\degree}$. The lower elevations may correspond to links to OGS that have sub-optimal positioning for the orbit under consideration. We also see that employing a $90^\circ$ (i.e., zenith) pass would only yield minor improvements in terms of generated SKL. For details on the reference scenario and the simulation methodology, see the next Sections.

\paragraph*{Custom inclination orbits}
Satellites in SSO have the advantage of servicing all locations on Earth, although polar regions are covered disproportionately often, as each satellite sweeps those regions around 15 times per day. Using, instead, custom inclination orbits, the regions that feature the highest coverage are located at latitudes equal to $i^\circ$\,N and $i^\circ$\,S, where $i$ is the orbit inclination. Analytical approximations of the average link coverage time, including the satellite height and minimum required elevation angle, can be obtained\cite{li2002analytical}. The inclination can then be optimised for better servicing the areas where the highest density of end-users is located. Furthermore, in this case, having an orbit period that is an exact multiple of Earth's day is not required, which gives more flexibility in the choice of orbit height. For instance, flying the satellite at a lower altitude $h=\SI{400}{\kilo\meter}$ allows improving the link budget for a zenith link by around $\SI{3}{dB}$, compared to the reference SSO.

Walker-type constellations can be constructed from several equally spaced orbital planes\cite{wertz1999space}. These are employed in current large satellite constellations used to provide satellite-based internet services, and similar configurations could be employed in commercial SatQKD implementations. An analysis of these is beyond the scope of the current work as it would require in-depth system performance evaluations for multi-user scenarios, potentially involving the use of QKD inter-satellite links.

\subsection{Reference parameters of the optical link budget}

\begin{figure*}[t]
\centerline{\includegraphics[width=\textwidth]{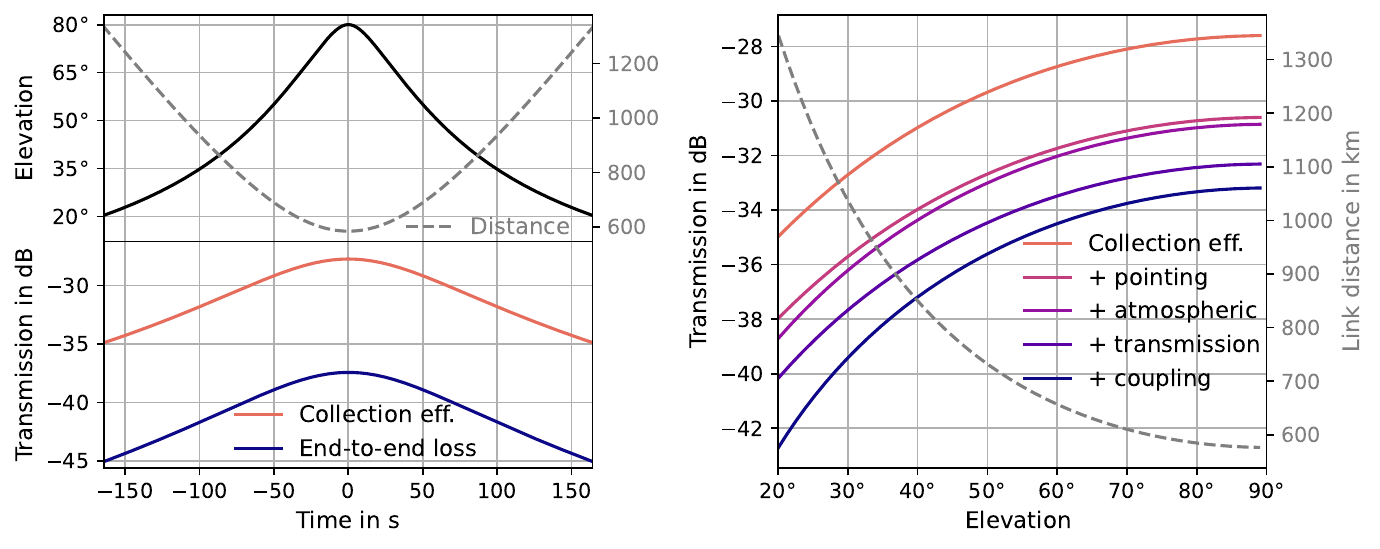}}
\caption{The upper left shows the elevation as well as the distance from the OGS over time for the reference orbit considered here. The lower left shows the resulting collection efficiency and end-to-end loss. On the right hand side, the different factors contributing to the total end-to-end loss over the elevation are presented.}
\label{fig:LinkBudget}
\end{figure*}

Here we provide a set of reference parameters for the performance of the satellite LCT, OGS telescope and associated subsystems. The given parameters are empirical, derived either form engineering requirements or from the specification parameters of real systems employed or under development at DLR. We thus deem these to be realistically achievable, provided that sufficient engineering optimisation effort is put in. These then allow us to compute a dynamical link budget for each point along the satellite pass. An example is presented in Figure~\ref{fig:LinkBudget}, where it is assumed that C-band quantum signals detected by the fibre-coupled SNSPD system presented in Table~\ref{tab:Detector_Specs}.

\paragraph*{LCT parameters}
We use as a reference satellite LCT system the one of QUBE-II\cite{Hutterer2022}. This will employ a telescope with $D_\Tx = \SI{85}{\milli\meter}$ diameter as primary aperture, designed for poly-chromatic operations. For the sake of comparison, we assume that similarly performing terminals can be manufactured for both the case where $\SI{1550}{\nano\meter}$ or $\SI{850}{\nano\meter}$ is employed as the quantum signal wavelength. In both cases we assume that a collimated Gaussian beam is expanded at the LCT external aperture; the beam waist is set to $w_0 = (D_\Tx/2)/\alpha$, where $\alpha = 1.12$ is the truncation ratio yielding highest antenna gain\cite{klein1974optical}. The $M^2$ parameter (an empirical multiplicative factor determining how divergent the beam is, compared to the diffraction-limited case) is set to $M^2=1.2$ for both wavelengths. This requires a better surface polishing quality for the shorter wavelength, but it is a sensible engineering target requirement in both cases. As discussed in Section~\ref{subsec:SPS_WCP} for QKD protocols employing WCP, transmission losses within the QKD transmitter terminal, as well as beam truncation losses, are irrelevant, since they can be compensated by increasing the optical power to reach a target signal intensity at the external aperture. This results in an antenna gain of \SI{102.2}{dB} for $\SI{1550}{\nano\meter}$ and \SI{107.5}{dB} for $\SI{850}{\nano\meter}$.

The presence of pointing jitter due to satellite platform vibrations will result in a decrease of the average signal power received by the OGS. The exact value will depend on the performance of the specific PAT employed. In the present analysis we assumed a fixed value of \SI{3}{dB} to model the power decrease due to pointing loss. This is employed for both wavelengths even though in the $\SI{850}{\nano\meter}$ scenario it will require higher engineering efforts to be reached due to the smaller beam divergence. 

\paragraph*{Optical channel parameters}
We have employed the MODTRAN software to simulate atmospheric absorption and scattering at the reference wavelengths as a function of the elevation angle. The assumed visibility was set to \SI{23}{\kilo\meter}. The free-space loss, as defined in Friis equation for the link budget, is computed from the link range and signal wavelength\cite{klein1974optical}. For the estimation of the BGL we have only modelled the diffuse sky radiance from moonlight, as it is typically the largest contribution in night-time conditions. In presence of a full-Moon the sky radiance reaches around $\SI{4}{\micro\watt \per {cm^2} \per \steradian \per nm}$ at $\SI{850}{\nano\meter}$ and around $\SI{1}{\micro\watt \per {cm^2} \per \steradian \per nm}$ at $\SI{1550}{\nano\meter}$. These will result in an increase of the QBER, depending on the receiver system parameters.

\paragraph*{OGS parameters}
We use the telescope on the rooftop of the DLR Institute of Communications and Navigation as reference OGS system; it consists of a two-axes mounted Cassegrain telescope featuring an $\SI{800}{\milli\meter}$ diameter primary mirror and a $\SI{300}{\milli\meter}$ diameter secondary mirror. We have assumed \SI{1.0}{dB} of optical losses in the free-space path (e.g. due to finite mirror reflectivity) for both wavelengths. As previously discussed, an OGS may be employed either as a fiber-coupled system or as a free-space-coupled one. In the latter case we assume that, by using large area detectors, the coupling losses can be made negligible. In the former case a high-order AO is required to achieve good focusing into the optical fibre. Even though effective models predicting the performance of AO system do exist\cite{scriminich2022optimal}, they do not capture the complete physics of the device. Therefore, we opted here to conservatively employ a fixed value of \SI{5.0}{dB}, as it is expected to be a sensible system engineering requirement for all elevations.

Background-light can be coupled into the detector and thus result in an increase of spurious photon detections, which may increase the QBER and shall therefore be assessed. Fibre-coupled systems naturally feature a very strong spatial filtering of the BGL, as matching to the transmitted SMF mode is required, having an effective area of only a few square micrometers. We compute the expected BGL counts at \SI{1550}{\nano\meter} for the SNSPD system in Table~\ref{tab:Detector_Specs}, assuming an effective spectral filter bandwidth of $\SI{5}{\nano\meter}$ and a optimal coupling factor $\beta = 1.12$\cite{BGL}. This results in around $\SI{8}{click\per\second}$ for full-Moon illumination conditions. For free-space detectors the coupling can be determined using a geometric optics approach. We assume that a circular field-stop with \SI{25}{\micro\meter} diameter is employed, restricting the detector area that can be illuminated, and that the system's effective focal length is $\SI{2}{\meter}$, resulting in a field-of-view half-angle of $\SI{6.25}{\micro\radian}$. The resulting backgroundlight rates for the detectors referenced in Table~\ref{tab:Detector_Specs} read $\SI{33}{click\per\second}$ (ID Qube) and $\SI{380}{click\per\second}$ (SPDCM-850-14), the difference stemming from the different BGL rates at the two wavelengths as well as the different detector efficiencies.

\subsection{Physical encoding of the quantum information}

In the previous section, it was already argued that at the current state of technology, implementations of the decoy-state BB84 protocol will most likely outperform any other QKD protocol in terms of implementation security and system performance. However, even within this class of protocols different realizations can be considered. One such choice is the physical realization employed to encode a qubit. This requires two orthogonal (i.e., fully distinguishable) optical modes, which may be based on the polarisation, temporal, or spatial degrees of freedom. Here, we will limit our considerations to the two most common choices of encoding, namely polarisation and time-bin. While it is in principle possible to encode qubits (and also their high-dimensional generalisations) in the orbital angular momentum of a propagating beam, the complexity of such implementations currently limit the maximum FSO communication distance to a few kilometers\cite{cozzolino2019high}.

\paragraph*{Polarisation encoding}
In polarisation encoding, Alice and Bob encode their bit values into two orthogonal polarisation modes of photons. These may be linearly or circularly polarised states. The convention we adopt here is to identify horizontal/vertical polarisations with the Pauli-Z basis, diagonal/antidiagonal polarisations with the Pauli-X basis, and left/right-handed polarisations with the Pauli-Y basis. Here we consider the use of linearly polarised light, as it is the choice employed most often. The use of the Pauli-Y basis has an advantage in SatQKD links: since the identification of left- and right-handed is invariant under rotation of the terminals around their line-of-sight (but is only affected by atmospheric birefringence, which is negligible), an active reference-frame tracking system for state discrimination in this basis. 

\paragraph*{Time-bin encoding encoding}
In time-bin encoding, the information is encoded in the temporal mode of the photon, with the early/late time-bins corresponding to the Pauli-Z basis and employing the relative phase between the two temporal modes to encode the Pauli-X basis. This has the advantage that the misalignment between Alice and Bob in the Z-basis (i.e. the overlap between early and late signals at the receiver) is usually negligible, allowing for a small QBER in this basis. On the other hand, due to the time-bin encoding each qubit governs at least two time slots, which might then require higher engineering efforts to achieve the same qubit rate as compared to polarisation encoding. 

During a LEO satellite pass the radial velocity varies by several kilometers per second, leading to variations of the signal frequency received at the OGS due to the Doppler shift. This results in a varying visibility in the receiver interferometer: for typical implementations the visibility oscillates between $-1$ and $+1$ multiple times over the pass. Therefore, this effect has to be tracked and compensated, otherwise the average X-basis QBER will approach $50\%$ and no secure key can be generated. The Doppler shift compensation can be achieved by several means, a simple one consisting in tuning the laser frequency by adjusting the driving voltage.

\subsection{Parameter optimisation}
\label{sec:results}

Within the decoy-state version of BB84, the number of decoy states to employ needs to be considered. The free parameters to be optimised over are then the different decoy intensities, the corresponding probabilities of sending them, and the probability of Alice and Bob choosing the Z-basis.

\begin{figure*}[t]
\centerline{\includegraphics{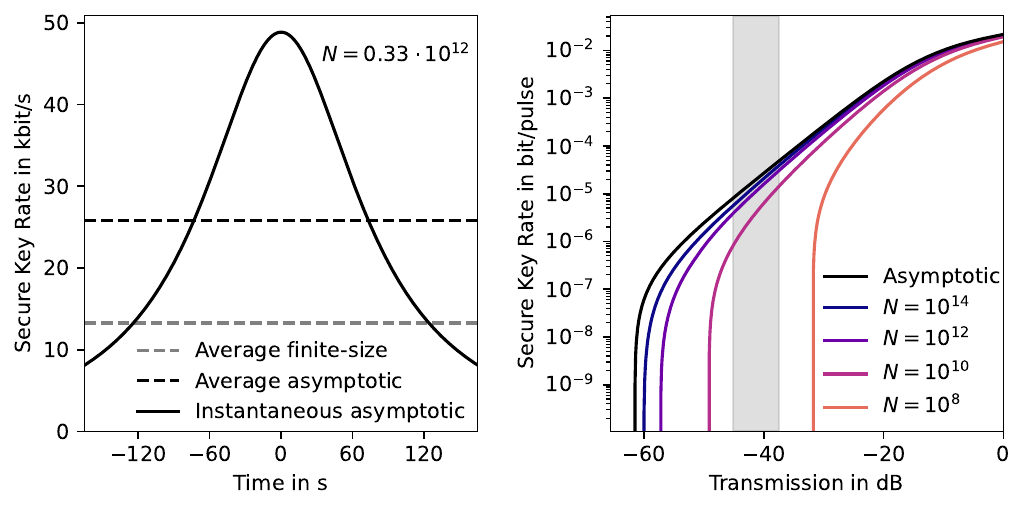}}
\caption{Left: instantaneous asymptotic SKR optimised at each point in time compared to the average SKR obtained over one satellite pass of the reference orbit. Right: impact of a finite amount of emitted pulses $N$ compared to the asymptotic limit of infinite data size, with the shaded region representing the range of transmissions covered by the reference orbit.}
\label{fig:AsymptoticFiniteSKR}
\end{figure*}

\paragraph*{Two decoy states}
The most common choice is the two-decoy state protocol, which features a signal intensity $\mu$ and two additional decoy intensities. It has been shown that the optimal value for the lowest decoy intensity is close to $0$; therefore, we call it the vacuum intensity, while the remaining decoy intensity $\nu$ fulfills $0 < \nu < \mu$\cite{Ma_2005}. 

In general, adding more and more decoy intensities should yield an increasingly better estimation of the parameters and hence result in a higher key rate. However, in practice, the (minor) improvements are outweighed by the increased system complexity. On the other hand, it has been argued that the rounds in which the vacuum intensity is emitted can never be used to establish information between Alice and Bob. Hence, it might be beneficial to omit the vacuum decoy state all together and estimate the amount of events with vacuum in the receiver input in a different way. 

\paragraph*{One decoy state}
By comparing the keys before and after error correction, Alice and Bob obtain knowledge about the QBER and thus the total number of erroneous events. For a worst-case estimate, they assume that all of these events resulted from vacuum at the receiver input. Such events should always yield a QBER of $\SI{50}{\percent}$, hence Alice and Bob obtain an upper limit for the number of vacuum events that can be used to estimate the amount of privacy amplification necessary to establish a secure key\cite{Rusca_2018_1decoy}.

\paragraph*{Secure key length}
To generate a secure key a sufficiently large number of signals have to be exchanged and accumulated, otherwise statistical fluctuations result in loose bounds on Eve's information, ultimately resulting in vanishing SKL. Depending on the system parameters, one may decide to accumulate signals over multiple satellite passes, in order to allow post-processing over longer blocks of bits. Alternatively, the key may be generated from the quantum signals exchanged over a single satellite pass; this allows for lower latency in the delivery of the secure key. Finally, one could instead subdivide the satellite pass in segments having similar QBER, allowing for a better parameter optimisation and higher SKR.

In contrast to fiber-based implementations, in SatQKD the channel is subject to varying transmission losses, due to scintillation and the change in relative position between satellite and OGS. Here, we choose to accumulated bits over one satellite pass of the reference orbit for joint post-processing. To perform QBER estimation, in the security proofs it is required that the system parameters (i.e.\ decoy intensities, their probabilities, and the Z-basis selection probability) are constant. Therefore, these parameters need to be optimised for the whole pass, instead of being subject to optimisation for each point in time. The left-hand side of Figure~\ref{fig:AsymptoticFiniteSKR} shows a comparison of the asymptotic SKR that would be achieved by such point-wise optimisation, compared to the result obtained by finite-size block post-processing over one pass.

The amount of secure key bits generated from one satellite pass then reads
\begin{equation}
    \ell = s_{\upZ,0} + s_{\upZ,1} \left[ 1 - h\left(\mathcal{E}_{\upZ|\upX,1}\right) \right] - \lambda_{\mathrm{EC}} - a \log_2(b/\varepsilon_{\mathrm{sec}}) - \log_2(2/\varepsilon_{\mathrm{corr}}) \,,
\end{equation}
where $a=6$ and $b=19$ or $b=21$ for the one- and two-decoy state implementation, respectively \cite{Rusca_2018_1decoy}. Here, $s_{\upZ,0}$ and $s_{\upZ,1}$ are (lower bounds to) the amount of detections given zero or one photon at the receiver input, respectively, $\mathcal{E}_{\upZ|\upX,1}$ is (an upper bound to) the average QBER of single photons measured in the Z-basis if they had been measured in the X-basis instead and $\lambda_{\mathrm{EC}}$ describes the amount of bits revealed during error correction.

Signals exchanged at lower elevations usually exhibit larger QBER, such that their inclusion can even reduce the length of the generated secure key\cite{sidhu2022finite}. Hence, specifically for satellite QKD systems, the minimum elevation considered in post processing is another parameter to be optimised. An overview of the optimal parameters and the resulting secure key length obtained over one satellite pass of the reference orbit for different detection systems, encoding schemes and numbers of decoy states is presented in Table~\ref{tab:Detector_KeyLength}.

\begin{table}
    \centering
    \begin{tabular*}{\textwidth}{@{\extracolsep\fill}llcccccccr}
        \toprule
        \multicolumn{1}{c}{\textbf{Detector}} & \multicolumn{1}{c}{\textbf{Encoding}} & \multicolumn{1}{c}{\textbf{\# Decoys}} & \multicolumn{6}{c}{\textbf{Optimal Parameters}} & \multicolumn{1}{c}{\textbf{SKL}}\\
        & & & min. elev. & $\mu$ & $\nu$ & $p_\mu$ & $p_\nu$ & $p_\upZ$ &\\
        \midrule
        \multirow{3}{*}{SNSPD ID281} & Polarisation & $2$ & $\SI{20}{\degree}$ & $0.67$ & $0.17$ & $0.82$ & $0.17$ & $0.90$ & $\SI{4.38}{\mega\bit}$\\
        & Polarisation & $1$ & $\SI{20}{\degree}$ & $0.58$ & $0.15$ & $0.81$ & $0.19$ & $0.88$ & $\SI{3.65}{\mega\bit}$\\
        & Time-bin & $2$ & $\SI{20}{\degree}$ & $0.80$ & $0.19$ & $0.81$ & $0.17$ & $0.88$ & $\SI{4.97}{\mega\bit}$\\
        \midrule
        \multirow{3}{*}{ID Qube NIR} & Polarisation & $2$ & $\SI{44}{\degree}$ & $0.55$ & $0.24$ & $0.67$ & $0.24$ & $0.73$ & $\SI{0.34}{\mega\bit}$\\
        & Polarisation & $1$ & $\SI{51}{\degree}$ & $0.52$ & $0.15$ & $0.72$ & $0.28$ & $0.67$ & $\SI{0.09}{\mega\bit}$\\
        & Time-bin & $2$ & $\SI{39}{\degree}$ & $0.65$ & $0.25$ & $0.68$ & $0.23$ & $0.73$ & $\SI{0.52}{\mega\bit}$\\
        \midrule
        \multirow{3}{*}{SPCM-850-14} & Polarisation & $2$ & $\SI{20}{\degree}$ & $0.69$ & $0.14$ & $0.89$ & $0.10$ & $0.93$ & $\SI{31.13}{\mega\bit}$\\
        & Polarisation & $1$ & $\SI{20}{\degree}$ & $0.60$ & $0.12$ & $0.89$ & $0.11$ & $0.92$ & $\SI{27.59}{\mega\bit}$\\
        & Time-bin & $2$ & $\SI{20}{\degree}$ & $0.84$ & $0.14$ & $0.88$ & $0.11$ & $0.92$ & $\SI{36.81}{\mega\bit}$\\
        \bottomrule
    \end{tabular*}
    \caption{Secure key length obtained by optimising the parameters over a single satellite pass for the different detection systems presented in Table~\ref{tab:Detector_Specs}. Note that this comparison assumes equal qubit generation rates, i.e. the three time-bins per qubit in time-bin encoding may result in a significant increase of the system requirements. The minimum elevation angle (min. elev.) was optimised over a range of $\SIrange{20}{80}{\degree}$ with a resolution of $\SI{1}{\degree}$.}
    \label{tab:Detector_KeyLength}
\end{table}

\section{Conclusions}
\label{sec:conclusions}

We have analyzed and discussed the central aspects of implementing satellite-based QKD systems with respect to their feasibility. We restricted the focus of our analysis to systems with a high TRL or ones that could realistically be deployed in the near-term future. Under this perspective, the most fitting architectural choice is to employ satellites in LEO running prepare-and-measure discrete-variable QKD protocols with quantum signals being sent only in the downlink direction. Within this architecture the satellite acts as a trusted node, which allows it to relay secure keys to pairs of end-users located anywhere on Earth as it orbits around it. This architecture is also considered in running and planned missions, e.g.\ in the SAGA first generation\cite{Lindman2023}. As patent from the current mission developments, our recommendation is not unorthodox; it must be noted, nonetheless, that no study has comprehensively justified it before.

We furthermore advise the use of decoy-state BB84 protocols as they currently feature the best trade-off between implementation security and system performance. Such protocols can be realized with different implementations: the quantum information can be physically encoded either in polarisation or in time-bins; the wavelength should be chosen in the bands around \SI{850}{\nano\meter} or \SI{1550}{\nano\meter} for which high performance single-photon detectors are readily available; the receiver systems themselves can be either in free space or fibre-coupled. None of the alternatives presented is definitely superior to the others, each of them offering different trade-offs to the designer of a SatQKD system.

\section*{Acknowledgments}
The work on which this report is based was done within the project QuNET funded by the German Federal Ministry of Education and Research under the funding code 16KIS1265 and the DLR project RoGloQuaN.The authors are responsible for the content of this publication.

\section*{Financial disclosure}
None reported.

\section*{Conflict of interest}
The authors declare no potential conflict of interests.

\bibliographystyle{unsrt}
\bibliography{1_Biblio}

\end{document}